\documentclass[fleqn,usenatbib]{mnras} 

\usepackage{float} 
\usepackage[dvipsnames]{xcolor}
\usepackage{ulem}
\usepackage{longtable,lscape}
\usepackage{caption} 
\usepackage{placeins}
\usepackage{adjustbox}
\usepackage{lastpage}   
\usepackage[T1]{fontenc} 
\usepackage{aecompl}	
\usepackage{graphicx} 
\usepackage{epsfig}
\graphicspath{{figs/}}
\DeclareGraphicsExtensions{.jpg, .png, .pdf, .eps}
\usepackage[multidot]{grffile}

\usepackage{mathtools}
\usepackage{amsmath}	
\usepackage{amssymb}	

\usepackage{xspace}

\newcommand{\msun}{~M$_{\odot}$\xspace}

\newcommand{\Lrad}{\ifmmode L_{\rm 1.4} \else $L_{\rm 1.4}$\fi\xspace}
\newcommand{\LHa}{$L_{\Ha}$}
\newcommand{\LHb}{$L_{\Hb}$}
\newcommand{\FHb}{$F_{\Hb}$}
\newcommand{\Loiii}{$L_{\oiii}$}

\newcommand{\densariv}{$n_{\rm e}$\ariv}
\newcommand{\denssii}{$n_{\rm e}$\sii}
\newcommand{\cHb}{$c({\Hb})$}

\newcommand{\hii}{\ifmmode \rm{H}\,\textsc{ii} \else H~{\sc ii}\fi\xspace}
\newcommand{\Ha}{\ifmmode {\rm H}\alpha \else H$\alpha$\fi\xspace}
\newcommand{\Hb}{\ifmmode {\rm H}\beta \else H$\beta$\fi\xspace}
\newcommand{\Hei}{He~{\sc i} $\lambda$5876}
\newcommand{\Heii}{He~{\sc ii} $\lambda$4686}
\newcommand{\Oiii}{[O~{\sc iii}] $\lambda$5007}

\newcommand{\Oii}{[O~{\sc ii}] $\lambda$3727}

\newcommand{\hei}{\ifmmode \rm{He}\,\textsc{i} \else He~{\sc i}\fi}
\newcommand{\heii}{\ifmmode \rm{He}\,\textsc{ii} \else He~{\sc ii}\fi}
\newcommand{\oi}{\ifmmode [\rm{O}\,\textsc{i}] \else [O~{\sc i}]\fi}
\newcommand{\oii}{\ifmmode [\rm{O}\,\textsc{ii}] \else [O~{\sc ii}]\fi}
\newcommand{\oiii}{\ifmmode [\rm{O}\,\textsc{iii}] \else [O~{\sc iii}]\fi}
\newcommand{\nii}{\ifmmode [\rm{N}\,\textsc{ii}] \else [N~{\sc ii}]\fi}
\newcommand{\sii}{\ifmmode [\rm{S}\,\textsc{ii}] \else [\ion{S}{ii}]\fi}
\newcommand{\ariv}{\ifmmode [\rm{Ar}\,\textsc{iv}] \else [\ion{Ar}{iv}]\fi}

\newcommand{\Hep}{\ensuremath{\mathrm{He}^{+}}}
\newcommand{\Hepp}{\ensuremath{\mathrm{He}^{++}}}

\title[Extragalactic planetary nebulae]
      {A study of extragalactic planetary nebulae populations based on spectroscopy.  I. Data compilation and first findings
}
\author[Delgado-Inglada et al.]
{G. Delgado-Inglada,$^{1}$\thanks{E-mail: gdelgado@astro.unam.mx}
 J. Garc\'ia-Rojas,$^{2,3}$\
 G. Stasi\'nska,$^{4}$
  J.S. Rechy-Garc\'ia$^{5}$ 
 \\
 $^{1}$Instituto de Astronom\'ia, Universidad Nacional Aut\'onoma de M\'exico, Apdo. Postal 70264, 04510, Ciudad de M\'exico, M\'exico.\\
 $^{2}$Instituto de Astrof\'sica de Canarias, E-38205 La Laguna, Tenerife, Spain\\
 $^{3}$Universidad de La Laguna, Dpto. Astrof\'isica, E-38206 La Laguna, Tenerife, Spain\\
 $^{4}$LUTH, Observatoire de Paris, PSL, CNRS 92190 Meudon, France \\
 $^{5}$Instituto de Radioastronom\'ia y Astrof\'isica, UNAM Campus Morelia, Apartado Postal 3-72, 58090 Morelia, Michoac\'an, México\\ 
}
\begin{document}

\maketitle

\begin{abstract} 
We compile published spectroscopic data and [\ion{O}{iii}] magnitudes of almost 500 extragalactic planetary nebulae (PNe) in 13 galaxies of various masses and morphological types. This is the first paper of a series that aims to analyze the PN populations and their progenitors in these galaxies. Although the samples are not complete or homogeneous we obtain some first findings through the comparison of a few intensity line ratios and nebular parameters. We find that the ionized masses and the luminosities in \Hb, \LHb, of around 30 objects previously identified as PNe indicate that they are most likely compact \hii regions. We find an anticorrelation between the electron densities and the ionized masses in M\,31, M\,33, and NGC\,300 which suggests that most of the PNe observed in these galaxies are probably ionization bounded. This trend is absent in LMC and SMC suggesting that many of their PNe are density bounded. The \Heii/\Hb values found in many PNe in LMC and some in M\,33 and SMC are higher than in the other galaxies. Photoionization models predict that these high values can only be reached in density bounded PNe. We also find that the brightest PNe in the sample are not necessarily the youngest since there is no correlation between electron densities and the \Hb luminosities. The strong correlation found between \LHb--\Loiii\ implies that the so far not understood cut off of the planetary luminosity function (PNLF) based on \oiii\ magnitudes can be investigated using \LHb, a parameter much easier to study. 
\end{abstract}



\section{Introduction}
\label{intro}
Planetary nebulae (PNe)  are produced at the end of the lives of low- and intermediate-mass stars when the stars leave the Asymptotic Giant Branch (AGB) after having lost most of their envelope  and increase their temperatures to the point when they can ionize the surrounding gas. The PNe central stars (CSPN) eventually evolve towards the white dwarf stage,  while the nebulae dissipate into the interstellar medium. The lifetimes of  PNe  $-$ i.e. the time where the nebulae are visible in the optical $-$ is dictated both by time during which  the central stars are luminous and able to ionize the surrounding medium and the expansion time of the nebulae. As known, the first is extremely dependent on the initial mass of the star \citep{paczynski71, schoenberner83, vassiliadis93, blocker95}. Recent evolution models for CSPN \citep[][with updated physical ingredients and atomic data]{millerbertolami16} predict higher luminosities and much shorter evolution times than found previously. This has important consequences on our understanding of the population of PNe in galaxies \citep{gesicki14, mendez17, valenzuela19}.

The best way to study the evolution of PNe is to consider them in external galaxies. Of course, their large distances makes their identification subject to errors, acquirement of their spectra difficult, and makes it impossible to measure their dimensions \citep[except in the Magellanic Clouds, see e.g.,][]{stanghellini03}.  But the fact that their distances are known (implying that their luminosities are known as well) and that their positions in the galaxies can be well determined makes their study priceless, in spite of the difficulties. The recent determinations of distances of PNe in the Milky Way using Gaia \citep{kimeswenger18, gonzalez19, stanghellini19, chornay20}, as well as studies of PNe in the direction of the Galactic bulge \citep{gorny04, Gorny2009} offer complementary approaches to the understanding of the evolution of PNe, but they are affected by other problems (difficulty to extract integrated values from the observations,  strong interstellar extinction in the case of Galactic bulge PNe, etc.).

The first comparisons of PN populations in different galaxies can be traced to \citet{webster76}. At that time data were available only for a few PNe in the Galactic bulge and the Magellanic Clouds. Twenty years later \citet{stasinska98} could consider 4 galaxies (M\,31, M\,32, LMC, and SMC), plus the Galactic bulge. These studies only scratched the surface of the issues worth studying. At present, the number of galaxies with spectroscopically observed PNe is much larger and allows the consideration of star-formation histories of the galaxies to understand their present-day PN populations. Other related works are those by \citet{richer93, richer06, richer11} and \citet{richer16}.

We have undertaken a compilation of published data on extragalactic PNe, to provide a general panorama for upcoming studies. Note that the objects chosen for spectroscopy are generally among the brightest, but they are not systematically the brightest ones in a given galaxy. So care must be taken not to over-interpret the data in terms of luminosity function.

In Section~\ref{requirements} we explain how we selected the data and why. In Section~\ref{data} we present the observational characteristics of the compiled sample, the line strengths, and the luminosties. In Section~\ref{uncert} we explain how we deal with uncertainties and we evaluate them. In Section~\ref{nature} we investigate whether the objects are real PNe or not. In Section~\ref{plots} we study different evolutionary aspects of the PNe. Finally, our conclusions are summarised in Section \ref{conclusions}. 


\section{Requirements for the samples}
\label{requirements}

\subsection{Galaxy selection}
\label{selection}

To construct an homogeneous set of observations we searched in the literature for extragalactic PNe spectroscopic data fulfilling several requirements that we explain in detail in the following:

\begin{itemize}
  \item We have selected galaxies with available spectroscopic data fulfilling the criteria summarized below for a minimum of 5 PNe. 
  \item When various sources are available for the same galaxy, we have discarded a few papers with observations of only a few PNe or with observations of worse quality than the other available sources.
  \item The available data must allow an estimation of the reddening (generally from the observed \Ha/\Hb ratio).
  \item  Intensities (or upper limits) must be available \textit{at least }for \Oii, \Hei, \Heii, \Hb, and \Oiii. This choice  is inspired by the work of \citet{stasinska98} where the authors were able to find differences between PNe in different galaxies from these few bright lines. Here we also used other lines from the list presented in Section \ref{line_selection}, when available.
  \item Since one of the objectives of this study is to understand the evolution of PNe by examining the distribution of certain line ratios as a function of luminosity, we only use observations of objects for which photometric estimation of the \Hb or [O~{\sc iii}] luminosity of the \textit{entire} object is available, either from calibrated photometry, or from spectrophotometry in the case where the PN is entirely covered by the spectroscopic aperture.
\end{itemize} 

These criteria exclude from our sample four dwarf galaxies: IC~10, NGC\,185, Sextans A, and Sextans B \citep{magrini05, magrini09, gonzalves12} because the number of PNe with the blue \oii\ lines is less than five in all of them. 

Overall we consider data from 13 galaxies. For each galaxy, we have collected information about morphological type, mass, star formation rate, metallicity ($Z$) and distance.

We have not included PNe data in NGC\,5128 from \citet{walsh15} because these authors reported a possible offset of the slits from the PN positions in their observations which might affect the conclusions derived from these data.

In Table \ref{tab:galaxies} we present some global parameters of the considered galaxies: 
\begin{itemize}
  \item col. 1: the galaxy name,
  \item col. 2: the morphological type according to NASA/IPAC Extragalactic Database (NED\footnote{The NASA/IPAC Extragalactic Database (NED) is funded by the National Aeronautics and Space Administration and operated by the California Institute of Technology.}),
  \item col. 3: the morphological type according to the Third Reference Catalogue of Bright Galaxies  \citep[RC3,][]{devaucouleurs91},
  \item col. 4: the total $B-V$ color index, $(B-V)_T^0$, corrected for galactic and internal extinction, and for redshift according to RC3,
  \item col. 5: the distance, as given by \citet{jarrett19}, taking the median of redshift-independent distances (Cepheid, Tip of the Red Giant Branch, Tully-Fisher, etc.) as tabulated in NED,
  \item col. 6: the distance estimated in the papers that studied the PNLF in these galaxies \citep{jacoby90, magrini01, Mendez01, ciardullo04, Corradi05, leisy05, merrett06, pena07b},
  \item col. 7: the extinction due to the Milky Way   from \citet{schlafly11},
  \item col. 8: the global stellar mass of the galaxy $M_{g}$ as obtained by \citet{jarrett19},
  \item col. 9: the total star formation rate as estimated by \citet{jarrett19} from the Wide-field Infrared Survey Explorer \citep[WISE,][]{wright10}, W3 band, SFR$_{W3}$,
   \item col. 10: the total star formation rate as estimated  from the WISE  W4, band SFR$_{W4}$,
   \item col. 11: the specific star formation rate as estimated by \citet{jarrett19}. 
\end{itemize}

\begin{figure*}
\centering
\includegraphics[width=\columnwidth, trim=40 0 60 0cm]{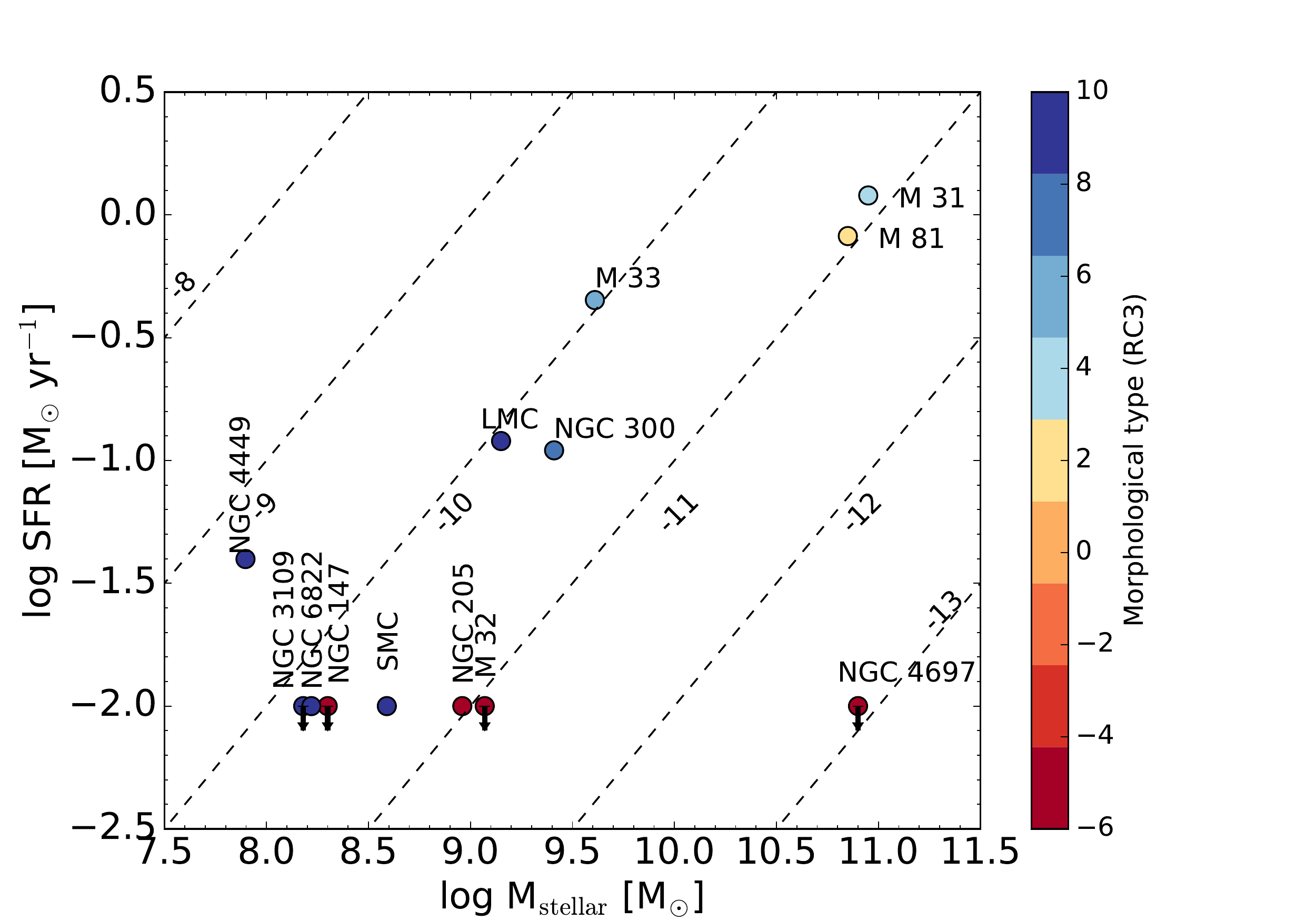}
\includegraphics[width=\columnwidth, trim=20 0 80 0cm]{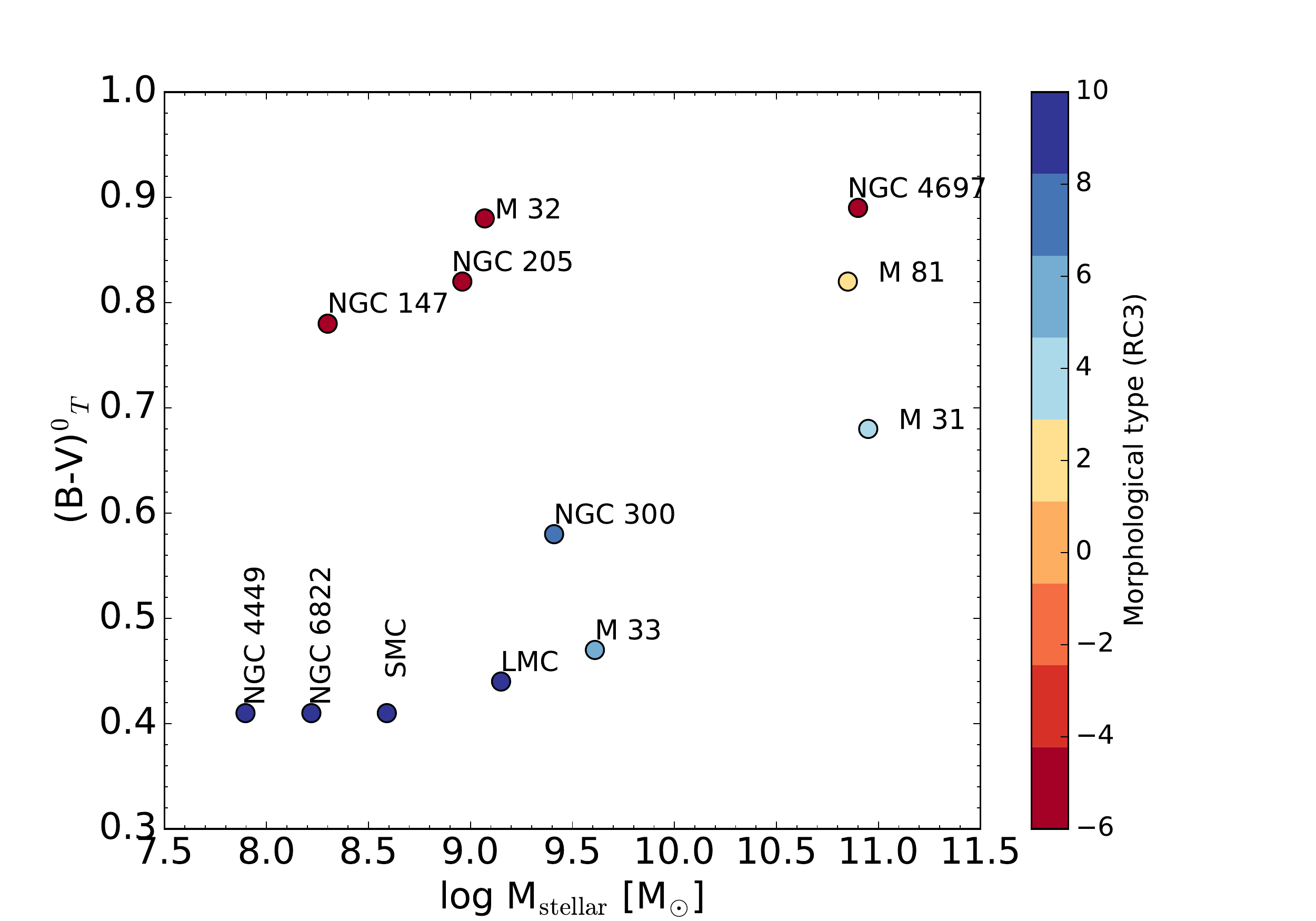}
\caption{Star formation rate (as estimated from the W3 band) and the total B$-$V color index (corrected for galactic and internal extinction and for redshift) as a function of the global stellar mass of the galaxy for the considered galaxies. Values taken from \citet{jarrett19}. In the left panel lines of constant specific star formation rate are plotted (dashed lines). The colorbar represents the morphological type according to RC3 \citep{devaucouleurs91}.}
\label{fig:sfrmgal}
\end{figure*}

\begin{table*}
\caption{The galaxies considered in this paper}
\label{tab:galaxies}
\begin{center}
\begin{tabular}{llccccccccc}
\hline
(1)         & (2)       & (3)       & (4)       & (5)       & (6)       & (7)       & (8)        & (9)  & (10) &  (11)\\
name        & morph type& morph type& $(B-V)_T^0$&distance$^a$  & distance$^b$ & $A_V$ & log $M_g$ & SFR$_{W3}$ & SFR$_{W4}$ & log sSFR \\
            & (NED)     & (RC3)     & (RC3)      & [Mpc]   & [Mpc] & [mag] 	& [\msun]       &[\msun yr$^{-1}$]&[\msun yr$^{-1}$] & [ yr$^{-1}$] \\
\hline                                          
M\,32        & E2        & -6        & 0.88       & 0.82   & 0.71   &   0.170   &  9.07     &$<$0.01    &$<$0.01  & -11.07 \\ 
NGC\,4697    & E2        & -5        & 0.89       & 10.50  & 10.50  &   0.081   & 10.90     &$<$0.01    & 0.04    & $<$-11.67 \\ 
NGC\,205     & E5 pec    & -5        & 0.82       & 0.82   & 0.85   &   0.170   &  8.96     & 0.01      & 0.0      & -10.96   \\
NGC\,147     & E5 pec    & -5        & 0.78       & 0.73   & 0.66   &   0.473   &  8.30     & $<$0.01   & $<$0.01 &  $<$-9.69 \\
M\,81        & SA(s)ab   & 2         & 0.82       & 3.65   & 3.84   &   0.220   & 10.85     & 0.82      & 0.52     & -10.94 \\ 
M\,31        & SA(s)b    & 3         & 0.68       & 0.77   & 0.78   &   0.170   & 10.95     & 1.20      & 0.58     & -10.88\\ 
M\,33        & SA(s)cd   & 6         & 0.47       & 0.86   & 0.94   &   0.114   &  9.61     & 0.45      & 0.33     & -9.96 \\ 
NGC\,300     & SA(s)d    & 7         & 0.58       & 1.93   & 1.81   &   0.035   &  9.41     & 0.11      & 0.08     & -10.37\\ 
NGC\,3109    & SB(s)m$^c$& 9         & $\ldots$   & 1.25   & 1.30   &   0.183   &  8.18     &$<$0.01    &$<$0.0   &$<$-9.23\\ 
LMC         & SB(s)m    & 9         & 0.44        & 0.05   & 0.05   &   0.206   &  9.15     & 0.12      & 0.03     & -10.07\\ 
SMC         & SB(s)m pec& 9         & 0.41        & 0.06   & 0.06   &   0.101   &  8.59     & 0.01      & 0.01     & -10.59\\ 
NGC\,4449    & SA(s)d    & 10        & 0.41       & 4.11   & $\ldots$   &   0.053   &  7.90$^d$ & 0.04$^d$  & 0.12$^d$ & -9.30$^d$\\ 
NGC\,6822    & IB(s)m    & 10        & 0.41       & 0.47   & 0.48   &   0.646   &  8.22     & 0.01      & 0.01     & -10.22\\ 
\hline
\end{tabular}
\begin{description}
$^{\rm a}$ From \citet{jarrett19}. \\
$^{\rm b}$ From the PNLF papers (the references are provided in the text).\\
$^{\rm c}$ Edge-on. \\
$^{\rm d}$ Computed from WISE data using the procedure of \citet{jarrett19}.
\end{description}
\end{center}
\end{table*}
%

In Fig. \ref{fig:sfrmgal} we show some global properties of the selected galaxies. The left panel shows the present star-formation rate SFR$_{W3}$  as computed from the  WISE W3 band, as a function of the total stellar mass of the galaxies.  The dashed lines correspond to loci of equal specific star-formation rates.  
The right panel shows the  $(B-V)^{0}_{T}$ colour as a function of stellar mass. The colours of the symbols indicate the de Vaucouleur types. 
It is thus seen that the PNe discussed in this paper belong to galaxies that cover a wide range of masses (roughly from 10$^8$ to 10$^{11}$ \msun), colours and morphological types (from very early to very late). At the high mass-end, the specific star formation rates, which can be understood as a ratio of present to past star formation rates, goes from 10$^{-11.67}$ to 10$^{-9.96}$. The populations of the PN progenitor masses are thus expected to be very different, being more massive for M\,31 which is a galaxy that is still forming stars and less massive for NGC\,4697  which mostly contains evolved stars (however, PNe in the bulge of M\,31 are expected to be a population similar to that of NGC\,4697). The listed specific star formation rates for M\,32 and NGC\,147 are upper values, but from their morphological types and galaxy colours, the progenitors of their PN population are also expected to be of low masses. 

\subsection{PN characterization}
\label{characterization}

Most of objects appearing in the papers presenting spectroscopy of extragalactic PNe were simply taken from photometric catalogs where they were classified as PNe. The criteria that are commonly used in these catalogs to distinguish PNe from {\hii} regions are the following: 1) the object has to be point-like, 2) it shows emission in {\oiii}, 3) it does not show emission in the continuum (or it is very faint), 4) the {\Heii} line is present (this criterion is sufficient but not necessary), and 5) the {\oiii}/{\Hb} intensity ratio is larger than $\sim$3--4. It should be noted that low excitation PNe, which correspond to the earliest stages of PN evolution are underrepresented because the selection generally favours objects bright in \oiii.

Other criteria that are used less frequently are: 1) the B and V magnitudes to confirm some ambiguous cases \citep{ciardullo04}, 2) the value of I({\oiii})/I({\Ha}+{\nii}) to estimate contamination from {\hii} regions \citep{magrini00}, 3) the visual magnitudes of the central star \citep[that are fainter for PNe than for {\hii} regions,][]{pena07a}, 4) the {\Ha} or {\Hb} flux that is lower in PNe than for \hii regions \citep{pena07a}, and 5) the value of {\sii}/{\Ha} intensity ratio \citep[that it has to be lower than 0.3 to exclude supernova remnants,][]{hdezmtnez09}. 

Only a few authors used the spectroscopic data to reclassify an object as a PN, an {\hii} region, or a supernova remnant \citep[e.g.,][]{pena07a, stasinska13, roth18}.

\subsection{Line selection}
\label{line_selection}

We compiled data on a moderate subset of lines. We are mostly interested in data allowing to estimate the evolution state of the PNe and, in a future paper, in the abundances of most common elements (He, O, N, Ne, S, Ar). We also kept some lines which can serve to estimate the data quality. 

The 28 lines we considered are: {\Oii} (in a few cases the [O~{\sc ii}] $\lambda$3726,29 lines are resolved), [Ne~{\sc iii}] $\lambda$3869, H~{\sc i} $\lambda$4102, H~{\sc i} $\lambda$4341, [O~{\sc iii}] $\lambda$4363, He~{\sc i} $\lambda$4471, {\Heii}, [Ar~{\sc iv}] $\lambda$4711, [Ar~{\sc iv}] $\lambda$4740, H~{\sc i} $\lambda$4861, [O~{\sc iii}] $\lambda$4959, [O~{\sc iii}] $\lambda$5007, [Cl~{\sc iii}] $\lambda$5518, [Cl~{\sc iii}] $\lambda$5538, [N~{\sc ii}] $\lambda$5755, He~{\sc i} $\lambda$5876, [O~{\sc i}] $\lambda$6300, [S~{\sc iii}] $\lambda$6312, [N~{\sc ii}] $\lambda$6548, H~{\sc i} $\lambda$6563, [N~{\sc ii}] $\lambda$6584,  He~{\sc i} $\lambda$6678, [\ion{S}{ii}] $\lambda$6716, [\ion{S}{ii}] $\lambda$6731, [Ar~{\sc iii}] $\lambda$7136, [O~{\sc ii}] $\lambda$7320, [O~{\sc ii}] $\lambda$7330, and [S~{\sc iii}] $\lambda$9069.


\section{The data}
\label{data}

\subsection{Description of the observing runs}
\label{obsruns}

Table \ref{tab:pnobs} presents the observational characteristics of the compiled observing runs. In this table we have compiled the following information:
\begin{itemize}
  \item col. 1: the galaxy name (and if the PNe are located in the disk or bulge),
  \item col. 2: the telescope,
  \item col. 3: the instrument/mode/grism ,
  \item col. 4: the total exposure times (in s),
  \item col. 5: the covered wavelength range
  \item col. 6: the spectral resolution (FWHM) as reported in the original references. In some cases, only the dispersion in \AA  /pix was reported; in these cases we estimated the spectral resolution assuming a canonical sampling of $\sim$3.5 pixels,
   \item col. 7: the observing mode. The options are Multi object spectroscopy (MOS) with slitlets, MOS fiber-fed, longslit, and echelle spectroscopy,
   \item col. 8: the observing aperture (in arcsec). In the case of slits, we consider the slit width, in the case of fibers, we put the fiber diameter, 
   \item  col. 9: $N$ the number of PNe used in this work,
   \item  col. 10: the reference.
\end{itemize}

\begin{table*}
\caption{Summary of the considered PN observations}
\label{tab:pnobs}
\begin{center}
\begin{scriptsize}
\begin{tabular}{lllllccccc}
\hline
Galaxy       & Telescope  & Instrument/   & Exposure    & Wavelength & FWHM &Observing & Aperture &  $N$  & Ref.$^{\rm b}$ \\
             &            & Grism         & time (s)  & range (\AA) &  (\AA) &mode$^{\rm a}$ & (arcsec) &  &  \\
\hline                                         
M\,32 	& 3.6m CFHT & MOS/U900 & 10800 & 3727--4341 & $\sim$3.6 & MOS/sl & 1.0 & 9 & (1) \\
        &           & MOS/B600 & 1800 & 3727--5876  & $\sim$6.0 & & &  &     \\
        &           & MOS/O300 & 900 & 4686--10000 & $\sim$14.0  & & &  &     \\
\hline
NGC\,4697& 8.2m VLT & FORS1/300V & 90000  & 4450--8650 & 10.0 &MOS/sl & 1.0 & 14 & (2) \\
        & 10m Keck\,1 & LRIS-B/ 400/3400 & 19800 & 3100--5770 & 2.0 &  MOS/sl & 1.0  &    &     \\
\hline
NGC\,205 & 8m Gemini-N & GMOS/B600         & 7200 & 3600--6400   &  & MOS/sl & 1.0 & 8 & (3)\\
         &             & GMOS/R400+G5305   & 7200 & 6200--10500   & $\sim3$ & MOS/sl & 1.0 & 8 & (3)\\
\hline
NGC\,147 & 8m Gemini-N & GMOS/B600+G5303 & 5600 & 3500--6500 & $\sim3$ & MOS/sl   & 1   & 6    &   (4)\\
         &             & GMOS/R400+G5305 & 3600 & 6200--10500  & $\sim8$ & MOS/sl   & 1   & 6    &   (4)\\
\hline
M\,81 (disk)	& 6.5m MMT     & Hectospec & 28800  & 3600--9100 & 6.0&  MOS/ff & 1.5 & 19?? & (5) \\ 
\hline
M\,31 (bulge) 	& 3.6m CFHT & MOS/B600 & 8100, 1200 & 3727--5876 & $\sim$6.0 & MOS/sl & 1.0 & 29 & (1) \\
M\,31 (bulge and disk)	& 4m Kitt Peak & R-C Spec/316 & 23400 & 3700--7400 & 6.9 & MOS/sl & 2.0 & 15 & (6) \\
M\,31 (outer disk) 	& 3.5m ARC-APO & DIS & 3600--15600 & 3700--9600 & 8.0 & longslit & 1.5--2.0 & 16  & (7) \\
    	& 8m Gemini-N  & GMOS/B600 & 2400--7200 & 3600--6400 & 7.0 & longslit & 1.5 &  & \\
M\,31 (outer disk)	& 10.4m GTC & OSIRIS/1000B & 6360--7020 & 3700--7850 & 6.3 & longslit & 0.8 & 2 & (8) \\
M\,31 (outer regions)	& 10.4m GTC & OSIRIS/1000B & 7200--8100 & 3700--7850 & 6.3 & longslit & 0.8 & 9 & (9) \\
M\,31 (substructures) 	& 10.4m GTC & OSIRIS/1000B & 4800--9600 & 3700--7850 & 5.5 & longslit & 1.0 & 7 & (10) \\
M\,31 (outer halo,	& 10.4m GTC & OSIRIS/1000B & 4800--9600 & 3700--7850 & 5.5 & longslit & 1.0 & 10 & (11) \\
 substructures)      	&           & OSIRIS/1000R & 2400--3600 & & 6.4 & & &  &      \\
\hline
M\,33 (disk, halo)   & 6.5m MMT  & Hectospec & 14400 & 3600--9100 & 6.0 & MOS/ff & 1.5 & 96  & (12) \\
M\,33    & 8.2m Subaru & FOCAS/300B & 2$\times$5400 & 3700--6000 & 4.5 & MOS/sl & 1.2 & 16 & (13) \\
(central region)        &       & FOCAS/300R (2nd) & 900, 1200 & 3700--5750 & 4.5 &  &  &  &  \\
        &        & FOCAS/VPH650 & 2400, 3600 & 5300--7700 & 4.0 & &  &  &  \\
\hline
NGC\,300 & 8.2m VLT & FORS2/600B & 12600, 12000 & 3600--5100 & 4.5 & MOS/sl & 1.0 & 26 & (14) \\
(disk)         &          & FORS2/600RI & 11880, 5400 & 5000--7500 & 5.0 & &  &  &  \\
         &          & FORS2/300I & 7200, 1800 & 6500--9500 & 10.0 & &  &  &  \\
\hline
NGC\,3109 & 8.2m VLT & FORS1/600B & 5400, 3904 & 3700--5900 & 7.8 &MOS/sl & 1.7 & 8 & (15)  \\
           &          & FORS1/600V & 4300, 2700 & 4650--6800  & 7.8 &  &  &       \\
           & 6.5m Magellan & MIKE & 5400 & 3350--9400 & 0.1 & Echelle & 1.0  &  &   \\
NGC\,3109 & 6.5m Magellan & MIKE & 900--5400 & 3350--9400 & 0.1 & Echelle & 1.0  & 8 & (16)  \\
\hline
LMC 	& 3.58m NTT & EMMI/3 settings & 600--3600 & 3635--8400 &  2.0--11.0 & longslit & 1.0--1.5 & 2 & (17) \\
LMC 	& ESO 1.52m & B\&C & 300--1800 & 3200--7800 & 3--4 & longslit &  ? & 90$^{\rm c}$ & (18)$^{\rm d}$ \\
     	& MPG/ESO 2.2m & EFOSC &  300--1800 & 3200--7800 & 3--4  & longslit &  ? &  &  \\
    	& ESO 3.6m & EFOSC  & 300--1800 & 3200--7800 & 3--4 & longslit & ?  &  &  \\
    	& 3.58m NTT & EMMI & 300--1800 & 3200--7800 & 3--4 & longslit &  ? &  &  \\

\hline
SMC 	& 3.58m NTT & EMMI/2 settings & 600--3600 & 3635--7830 &  2.0--3.8 & longslit & 1.0--1.5 & 1 & (17) \\
SMC 	& ESO 1.52m & B\&C  & 300-1800 & 3200--7800 & 3--4 & longslit & ? & 30$^{\rm c}$ & (18)$^{\rm d}$ \\
     	& MPG/ESO 2.2m & EFOSC & 300--1800 & 3200--7800 & 3--4 & longslit & ?  &  &  \\
    	& ESO 3.6m & EFOSC & 300--1800 & 3200--7800 & 3--4 & longslit &?  &  &  \\
    	& 3.58m NTT & EMMI & 300--1800 & 3200--7800 & 3--4  & longslit &?  &  &  \\
SMC 	& 3.58m NTT & EMMI/2 settings & 1200--4800 & 3670--7900 & 0.6--1.2 & Echelle/longslit & 1.5 & 12 & (19) \\
\hline
NGC\,4449	& 8.4m LBT  & MODS/G400L-G670L & 37800 & 3200--10000 & 4.1--5.8 & MOS/sl & 1.0  & 5  & (20) \\
\hline
NGC\,6822 	& 3.6m CFHT & MOS/B600 & 9000 & 3700--6700 & $\sim$7--8 & MOS/sl & 1.0 & 7 & (21) \\
NGC\,6822 	& 8m Gemini-S  & GMOS/B600 & 4400 & 4000--6700 & $\sim$2.7 & MOS/sl & 1.0 & 11 & (22) \\
            &              & GMOS/R600 & 3600 & 4800--7500 & $\sim$1.7 & &  &    &      \\
            & 8.2m VLT     & FORS2/600B & 5400 & 3700--5100  & 4.5 &  MOS/sl & 1.0  &   &    \\
            &              & FORS2/600RI & 4500 &  5000--7500 & 5.0 &    &   &    \\
            & 8.2m VLT     & FORS2/600B & 4500 & 3700--5100  & 4.5 & longslit & 1.0  &   &    \\
            &              & FORS2/600RI & 3000 &  5000--7500 & 5.0 &     &   &   &    \\
NGC\,6822 	& 10.4m GTC & OSIRIS/1000B & 5400 & 3700--7850 & 7.5 &  longslit & 1.5 & 5 & (23) \\
\hline
\end{tabular}
\end{scriptsize}
\begin{description}
$^{\rm a}$ (MOS/sl) Multi Object Spectroscopy/Slitlets; (MOS/ff) Multi Object Spectroscopy/Fiber-fed \\
$^{\rm b}$ References: (1) \citet{richer99}; (2) \cite{mendez05}; (3) \cite{gonzalves14}; (4) \cite{gonzalves07}; (5) \cite{stanghellini10}; (6) \cite{jacoby99}; (7) \cite{kwitter12}; (8) \citet{balick13}; (9) \citet{corradi15}; (10) \citet{fang15}; (11) \citet{fang18}; (12) \citet{magrini09b}; (13) \citet{bresolin10}; (14) \citet{stasinska13}; (15) \citet{pena07a}; (16) \citet{floresduran17}; (17) \citet{tsamis03}; (18) \citet{leisy06}; (19) \citet{shaw10}; (20) \citet{annibali17}; (21) \citet{richer07}; (22) \citet{hdezmtnez09}; (23) \citet{garciarojas16} \\
$^{\rm c}$ Some of the line intensities of these objects come from archival data. \\
$^{\rm d}$ Exact wavelength coverage was not reported by \citet{leisy06}. In this table the reported range is the one representing the maximum coverage in their observations. The resolution is in \AA/pix. Slit widths were not reported by \citet{leisy06}.
\end{description}
\end{center}
\end{table*}


As it is clearly shown in Table~\ref{tab:pnobs}, the data have been taken either with large aperture telescopes (diameter $\geq$ 6m) or with smaller telescopes but obtaining data for a significant fraction of PNe population of each galaxy. 
Although our aim is to have a sample of PNe spectra with a relatively homogeneous quality, we bear in mind several drawbacks that could affect the final quality of the analyzed spectra.

\subsection{Line fluxes}
\label{fluxes}

We have decided to use the reddening-corrected intensities provided by the authors, as well as the extinction at \Hb derived by them. Most of the authors provided the logarithmic extinction parameter, \cHb, that can be obtained through the comparison of the observed and theoretical (case B) line ratios of two Balmer hydrogen lines. For example, when H$\alpha$ and H$\beta$ lines are used:

\begin{equation*}
\frac{I_{\rm theo}({\rm H}\alpha)}{I_{\rm theo}({\rm H}\beta)} = \frac{I_{\rm obs}({\rm H}\alpha)}{I_{\rm obs}({\rm H}\beta)}\times10^{c({\rm H}\beta)(f({\rm H}\alpha)-f({\rm H}\beta))},   
\end{equation*}
where $f$($\lambda$) is the adopted reddening law \citep[e.g.][]{cardelli89}. In a few cases, we have derived the value of \cHb\ from the $E(B-V)$ given by the authors \citep{richer99, richer07, annibali17}. The relation between these two quantities depends on the form of the extinction curve: \cHb = $0.4 R_V E(B-V)$. We have used the same R$_V$ values as those adopted by the authors.

This is not a fully consistent approach, since the procedures for dereddening vary, depending on the instrumental setup, on the quality of data (e.g. whether the H$\gamma$ and H$\delta$ intensities can be used with confidence) and on the adopted reddening law. But we did not see any advantage of recomputing the reddening, since in the wavelength range we are considering, the reddening laws adopted by the different authors are equivalent. There are two cases where we have computed the extinction coefficient and corrected the observed lines because the uncertainties in the reddening-corrected intensities were not provided by the authors: the PNe from \citet{stanghellini10} for M\,81 and the PNe from \citet{magrini09} for M\,33. In both cases we have adopted $R_V = 3.1$ and the extinction law by \citet{cardelli89}.

The sky subtraction is a critical step when dealing with extragalactic PNe spectra. The accuracy of this subtraction relies on the observing technique as well as on the characteristics of each galaxy. Data obtained with fibers are the least reliable, since the ``sky'' spectra is taken far from the PN loci. Note that, in our compilation, only the data from \citet{stanghellini10} and \citet{magrini09} for M\,81 and M\,33 were obtained with fibers. Data obtained from longslit or MOS/slitlets observations are more reliable if the sky is measured on both sides of the object.  However, the ``sky'' emission, which actually is dominated by the light from stellar populations encompassed by the observing aperture, may be far from uniform. As an example of the complexity of sky subtraction we refer to the case of PN 29 in the bulge of M\,31; for this PN, \cite{jacoby99} and \citet{richer99} obtained very different line intensities. Using Integral Field Unit (IFU) observations, \citet{roth04} showed that indeed the ``sky'' emission around PN 29 was not uniform. These authors pointed out the importance of a proper knowledge of the contribution of the continuum light of unresolved stars with a small angular separation from the target or from  emission-line spectra of \hii regions and diffuse nebulosities of the interstellar medium to the observed spectra of point-like sources, and demonstrated that this contribution can be reasonably well
determined with IFU spectroscopy when using full two-dimensional spatial information and point-spread function fitting techniques. Unfortunately, among the data collected for our paper, only those from  \citet{roth04} allow such a treatment, and the number of PNe  in that paper which comply the requirements presented in Sect. \ref{selection} is not sufficient. Another option would be to estimate the stellar contribution directly from the \textit{stellar} lines present in the spectrum of the object, as done for example by \citet{Kreckel17} for PNe in the galaxy NGC 628 that were observed with the Multi-unit Spectroscopic Explorer \citep[MUSE;][]{Bacon10}. Sadly, the wavelength range of MUSE is restricted to $\lambda > 4650$ \AA, so these data do not appear in our compilation. There is no doubt that the advent of similar instruments but working at shorter wavelengths, such as BlueMUSE \citep{richard19}, will allow considerable progress in spectroscopic studies of extragalactic PNe. For the time being, we have to live with what is available, and remember that some caution should be taken when interpreting the results obtained from the emission-line spectra of extragalactic PNe, especially in regions containing  old stellar populations.

Taking into account the relatively low resolution of the majority of the spectra in the sample, and that we are interested in the density diagnostic [Ar~{\sc iv}] $\lambda$4740/[Ar~{\sc iv}] $\lambda$4711, we have carefully corrected the [Ar~{\sc iv}] $\lambda$4711 emission line from the contribution of He~{\sc i} $\lambda$4713 emission. To do this we used the theoretical He~{\sc i} $\lambda$4713/He~{\sc i} $\lambda$5876 ratio for $n_e$=10$^3$ cm$^{-3}$ and $T_e$([O~{\sc iii}]) derived from the spectra (or $T_e$=10$^4$ K in case the auroral [O~{\sc iii}] $\lambda$4363 is not detected in the spectra). We consider this a sufficiently good approach given the small dependency of the He~{\sc i} $\lambda$4713/He~{\sc i} $\lambda$5876 intensity ratio with $n_e$ (less than 10\% in the 10$^3$-10$^5$ cm$^{-3}$ density range). We have performed this correction in the spectra of the PNe of the following galaxies: M\,31 \citep{kwitter12, balick13, corradi15}, M\,33 \citep{magrini09, bresolin10}, NGC\,300 \citep{stasinska13},  NGC\,3109 \citep{pena07a}, LMC \citep{leisy06}, SMC \citep{leisy06, shaw10}, NGC\,4449 \citep{annibali17} and NGC\,6822 \citep{richer07, hdezmtnez09}; all these spectra  have a spectral resolution greater than 2--3 \AA\ and a detected line at $\sim$4711 \AA. A more detailed discussion on this correction has been done in Section~\ref{nature}.

Lines of [O~{\sc ii}] $\lambda\lambda$3726+29 are summed in the tables even if observations allow one to separate the lines. The only PNe where this doublet is resolved are: one object in NGC\,3109 observed by \citet{pena07a} and the sample studied by \citet{floresduran17}, the PNe in the LMC and SMC reported by \citet{tsamis03}, and the SMC sample observed by \citet{shaw10}.

\subsection{Luminosities}
\label{luminosities}
  
When available, the [O~{\sc iii}] luminosity was computed from photometry, using the [O~{\sc iii}] $\lambda$5007 magnitudes, m(5007), and the formula from \citet{jacoby89} to convert them into fluxes. This is especially important for PNe in the Ma\-ge\-lla\-nic Clouds (MC), which are larger than the observing slits. For the MC, we thus use only the data for which the magnitude m(5007) is given in \citet{jacoby90}. 

For the other galaxies, if the [O~{\sc iii}] $\lambda$5007 magnitude was not provided in the papers listed in Table~\ref{tab:pnobs} we have compiled them from \citet{ciardullo89, merrett06} for M~31, \citet{ciardullo89} for M~32, \citet{ciardullo04} for M~33, \citet{magrini01} for M~81, \citet{Corradi05} for NGC~147 and NGC~205, \citet{pena12, roth18} for NGC~300, \citet{pena07b} for NGC~3109, \citet{Mendez01} for NGC~4697, and \citet{leisy05} for NGC~6822. 

If no photometry is available, we directly used the measured fluxes from spectrophotometric data. This is the case for the PNe in NGC\,4449.

The values of m(5007) for PNe in M\,81 as provided by \citet{magrini01} were corrected for the interstellar extinction by only considering the foreground Galactic extinction towards M\,81. We uncorrected the magnitudes using the values of A$_V$ and $E(B-V)$ given by \citet{magrini01} and then computed the [O~{\sc iii}] and \Hb luminosities in the same way as for the other PNe, i.e. using the extinction coefficients and line ratios obtained from spectrophotometric data given by \citet{stanghellini10}.
  
It is worth mentioning that in NGC\,300 \citet{roth18} found a systematic offset with a median of $\sim$0.67 mag between their computed magnitudes from deep MUSE data and those derived by \citet{pena12} for the PNe in common, being the value of m(5007) from \citet{pena12} higher than those computed by \cite{roth18}. These differences can be due to problems in flux calibration and/or slit losses and translates into differences of a factor of 2 in the fluxes obtained by both authors. \citet{roth18} found a much better agreement with the magnitudes reported by \citet{soffner96} for PNe in common. We have used the magnitudes from \citet{roth18} when available and corrected the magnitudes from \citet{pena12} by 0.67 mag otherwise.


\section{Dealing with uncertainties and upper limits}
\label{uncertup}

\subsection{Determination of the uncertainties in line strengths}
\label{uncert}

In principle, determining uncertainties in extinction-corrected line strengths requires to start from raw, flux-calibrated data, estimate the extinction with its errors bars, then compute the uncertainties in extinction-corrected line fluxes. The data available in most of the references we used do not allow such a procedure.
 
We have followed slightly different approaches, depending on the information given in the original data source and on the availability of similar observations for a given galaxy.

If error bars are specified explicitly in the original source, we simply adopt them. This is the case for the PNe taken from \citet{richer99, gonzalves07, richer07, hdezmtnez09, magrini09, stanghellini10, stasinska13, garciarojas16, annibali17, floresduran17}. 

Sometimes no error is given for the \Hb intensity so we have estimated it from a logarithm linear fit between the relative uncertainties vs. the line intensities of the other lines. The PN samples for which we have to do this are those from \citet{bresolin10, kwitter12, balick13, fang15, fang18}.  

For PNe data of LMC and SMC from \citet{tsamis03} and \citet{shaw10} the authors assigned percentage uncertainties depending on the brightness of the lines. In this case we have attributed errors using a formula to reproduce the assigned uncertainties. For \citet{tsamis03} the formula is $\Delta$I/I $\sim$0.05/[I($\lambda$)/I(\Hb)]$^{0.4}$ , and for \citet{shaw10} it is $\Delta$I/I $\sim$0.08/[I($\lambda$)/I(\Hb)]$^{0.3}$. As for the PNe from \cite{leisy06}, these authors reported that the uncertainty of lines with intensities between 1--5\% I(\Hb) scales approximately with the inverse square root of their intensity ratio to \Hb. Accordingly, we decided to use the following formula to compute uncertainties: $\Delta$I/I $\sim$0.08/[I($\lambda$)/I(\Hb)]$^{0.3}$. Whenever possible, we checked the error bars attributed by us to the observations by looking at the published spectra. This of course is a very rough estimation, since for a given observing run the uncertainties depend on the line fluxes (and not on the ratio of the line flux with \Hb), on the wavelength (blue is always noisiest), and on the exposure time. Additionally, they should be estimated on observed data (not dereddened) but this makes little difference. 

The papers by \citet{corradi15, jacoby99, mendez05} do not provide uncertainties at all. For the M\,31 PNe studied by \citet{corradi15}, we have applied a logarithmic linear fit to the relative uncertainties vs. the flux of the line for the objects reported by \citet{balick13} and \citet{kwitter12}. We took advantage of the fact that both observations were made with the same telescope+instrument (OSIRIS+GTC) and similar configurations. Then we used the fit to estimate the uncertainties associated with the line fluxes reported by \citet{corradi15}. The M\,31 PNe studied by \citet{jacoby99} were observed with a different telescope and configuration than any other sample considered in this paper. Therefore, we decided to perform  a similar analysis to that carried out for the data from \citet{corradi15}. For the observations reported by \citet{mendez05} in NGC\,4697 we used the uncertainties reported by \citet{stasinska13} for PNe in NGC\,300 and a similar approach to the one that was used for M\,31. Both authors used similar telescope and instrumentation: FORS1/VLT MOS with grism 300V by \cite{mendez05} and FORS2/VLT MOS with grisms 600B, 600RI and 300I by \citet{stasinska13}. 

As for NGC\,3109, \citet{pena07a} provide relative uncertainties for a few ranges in the values of the line intensities. However, we decided to apply the same procedure as described above to derive the uncertainties for all the lines using the uncertainties reported by \citet{stasinska13} for PNe in NGC\,300 since the observations have similar characteristics in both galaxies. 

\begin{figure}
  \centering
  \includegraphics[width=\columnwidth, trim= 20 0 40 0cm]{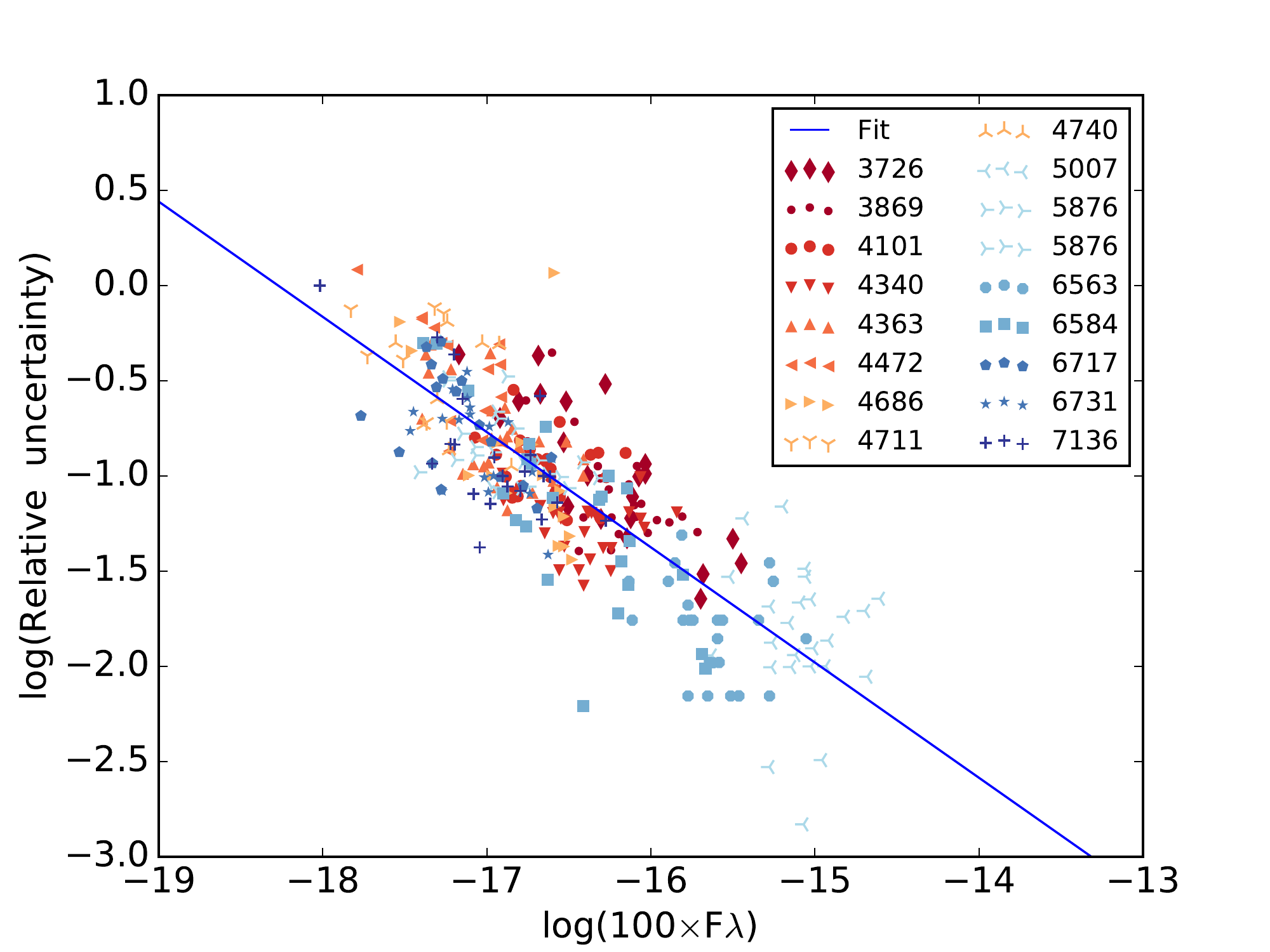}
  \caption{Values of the relative uncertainties as a function of the line intensities weighted by the total \Hb flux for different lines taken from \citet{stasinska13}. The solid line represents our fit.}
  \label{fig:fit}
\end{figure}

Figure~\ref{fig:fit} shows one example of the fits we have performed to estimate the uncertainties in the cases where the authors do not provide them.

The uncertainties in line ratios that are used later in the text,  such as \oiii/\Hb for example,  have been estimated from the uncertainties in \oiii\ and \Hb fluxes, propagated analytically, assuming the errors to be independent for simplicity.

Note that neglecting the uncertainty in the reddening correction has only a minor impact in optical line ratios, since most of  our objects have a value of \cHb\ much smaller than 1 (see Fig.~\ref{fig:lumdensext3}). The line ratio with the strongest impact is that of \Oii/\Hb, for which we estimate a possible error of about 25\% in the worst cases (i.e. when the relative uncertainties in \Ha and \Hb lines are $\sim$30\% and \cHb\ is of the order of 1, which occurs only in a small minority of cases).

\subsection{Upper limits}
\label{upperl}
 
In general, we consider upper limits as stated in the original papers. If upper limits of any of the necessary lines ([\ion{O}{ii}] $\lambda$3727, [\ion{O}{iii}] $\lambda$5007, \ion{He}{i} $\lambda$5876, and \heii\ $\lambda$4686) have not been reported in the original references, we have adopted as an upper limit the uncertainty of the faintest  line reported in the same wavelength range.

\subsection{Graphical evaluation of uncertainties}
\label{graphuncert}
 
Figures appearing hereafter are presented as `small multiple diagrams' according to the denomination of \citet{tufte90} to give a panoramic view of our sample. In case one is interested in details, the figures can be enhanced, since they are published at full resolution.

\begin{figure*}
\flushleft
\includegraphics[width=\textwidth, trim= 20 0 40 0cm]{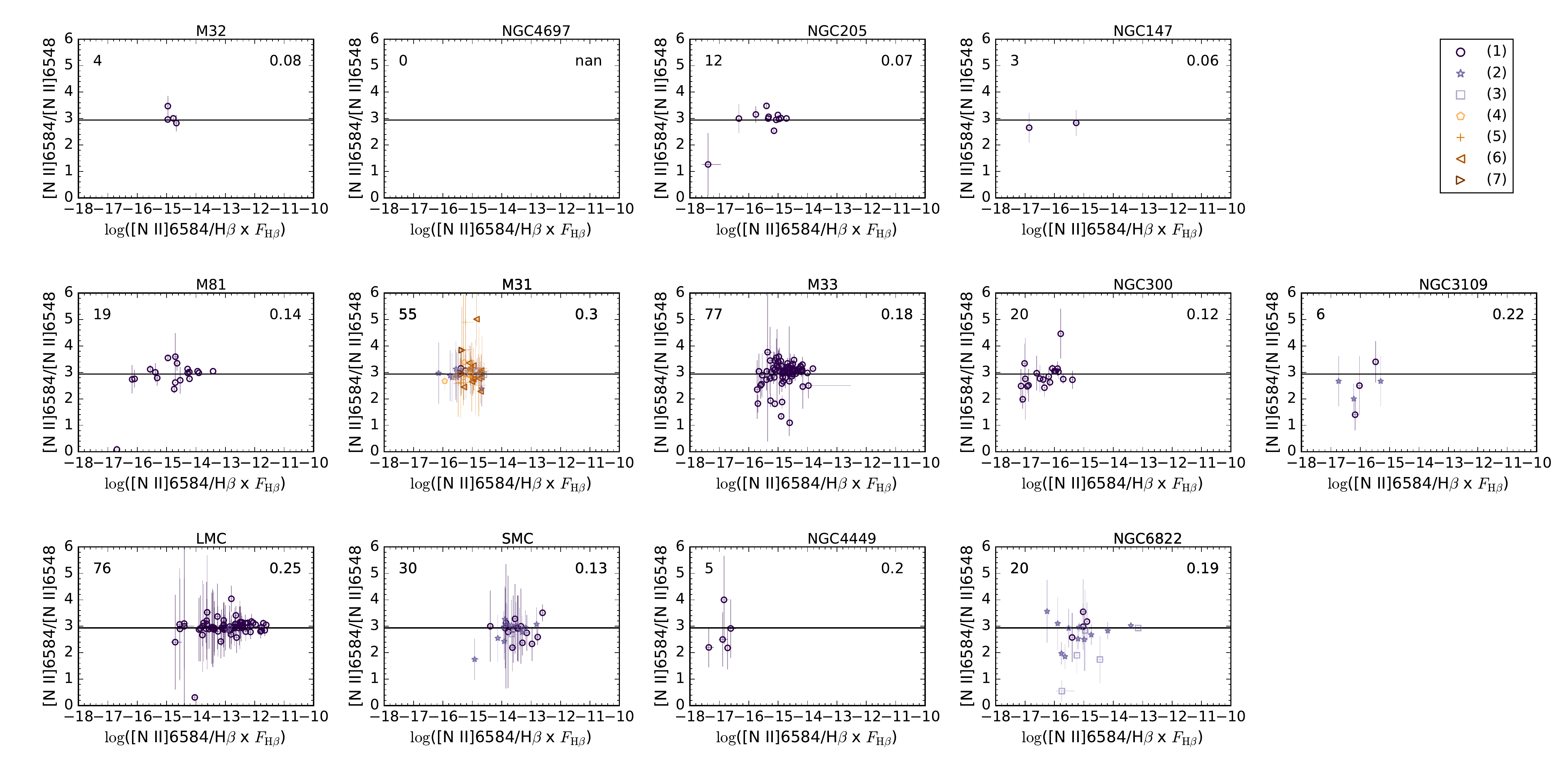}
\caption{Assessment of accuracies in line fluxes. Values of [\ion{N}{ii}] $\lambda$6584/$\lambda$6548 ratio as a function of [N~{\sc ii}] $\lambda$6584/H$\beta\times$\FHb. \textbf{Upper panels:}  early type galaxies, \textbf{middle panels:}  spiral galaxies, \textbf{lower panels:} irregular galaxies. The number that appears at the top left corner of each panel indicates the number of PNe plotted in the panel. The number that appears at the top right corner of each panel indicates the weighted standard deviation of the [\ion{N}{ii}] $\lambda$6584/$\lambda$6548 values divided by its theoretical value. The color- and symbol-code for the galaxies with more than one reference is based on the following numbering: M\,31: (1) B13, (2) C15, (3) F15, (4) F18, (5) J99, (6) K12, (7) R99. M\,33: (1) M09, (2) B10. NGC\,3109: (1) FD17, (2) P07, LMC: (1) LD06, (2) T03. SMC: (1) LD06, (2) S10, (3) T03. NGC\,6822: (1) GR16, (2) HM09, (3) R07.}
\label{fig:quality1}
\end{figure*}

\begin{figure*}
\flushleft
\includegraphics[width=\textwidth, trim= 20 0 40 0cm]{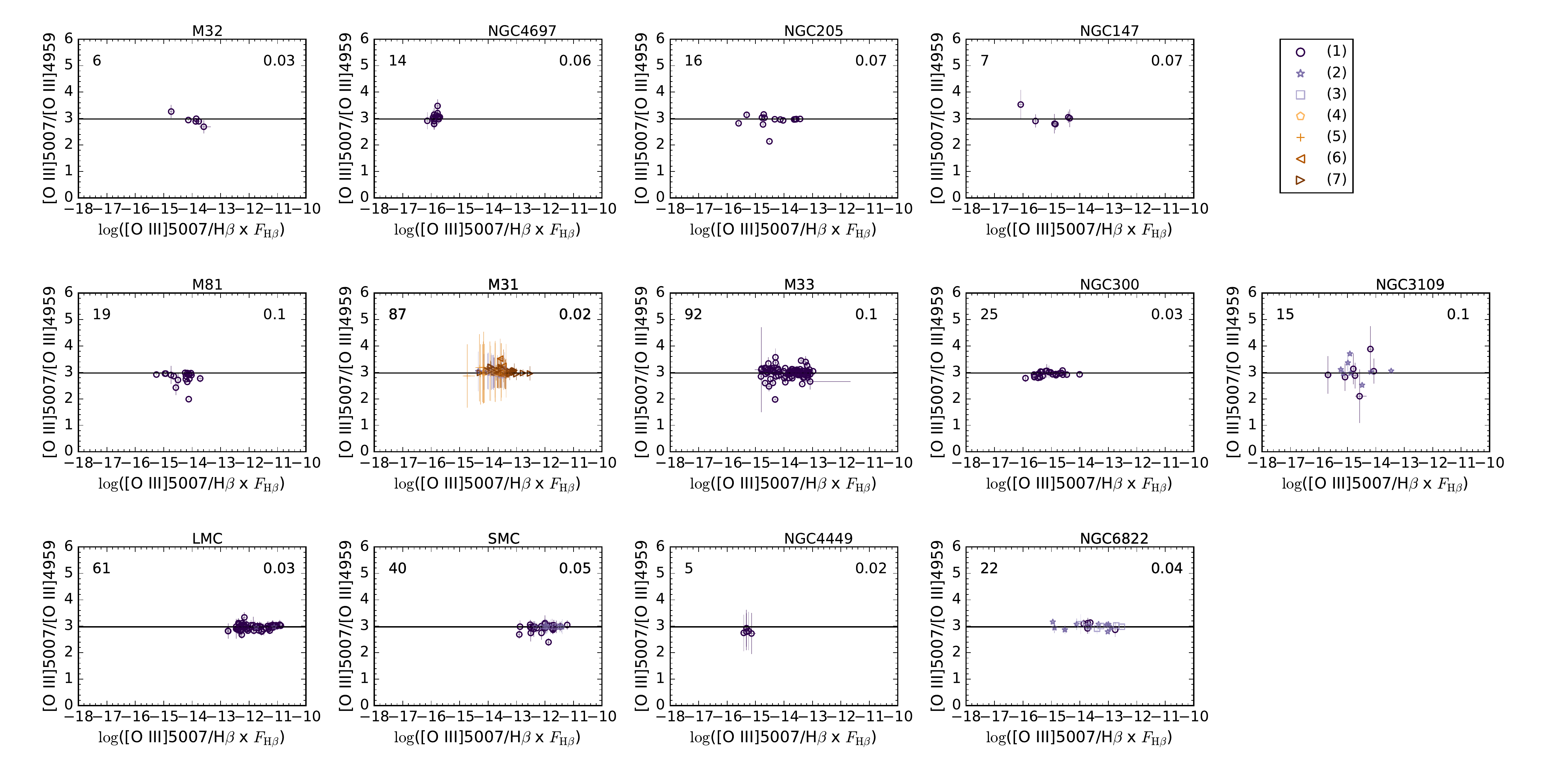}
\caption{Assessment of accuracies in line fluxes. Values of [O~{\sc iii}] $\lambda$5007/$\lambda$4949 ratio as a function of [O~{\sc iii}] $\lambda$5007/H$\beta\times$\FHb. \textbf{Upper panels:}  early type galaxies, \textbf{middle panels:}  spiral galaxies, \textbf{lower panels:} irregular galaxies. The number that appears at the top left corner of each panel indicates the number of PNe plotted in the panel. The number that appears at the top right corner of each panel indicates the weighted standard deviation of the [O~{\sc iii}] $\lambda$5007/$\lambda$4949 values divided by its theoretical value. The numbering of the references is the same as in Fig.~\ref{fig:quality1}.}
  \label{fig:quality2}
\end{figure*}

\begin{figure*}
\flushleft
\includegraphics[width=\textwidth, trim= 20 0 40 0cm]{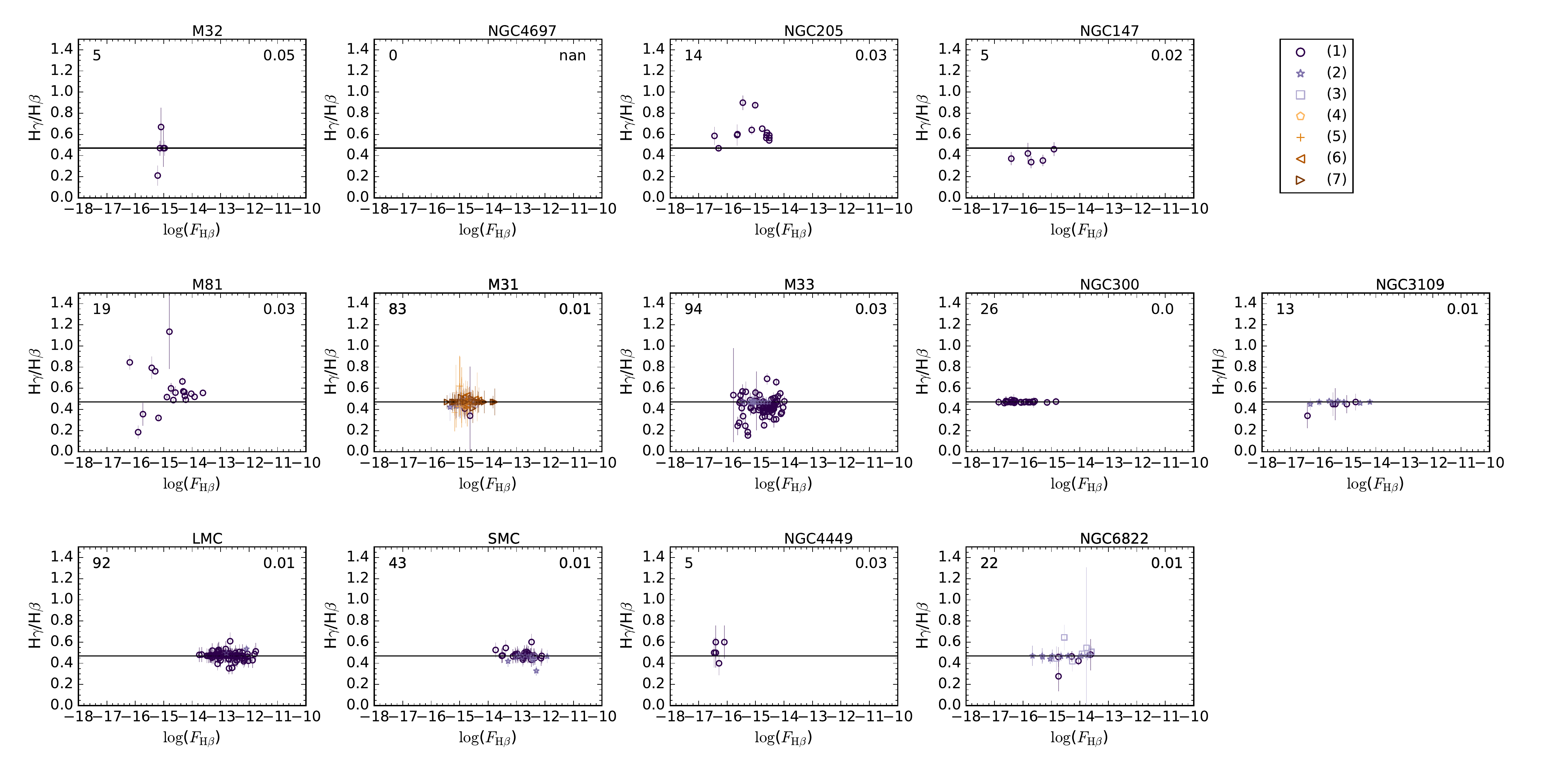}
\caption{Assessment of accuracies in line fluxes. Values of H$\gamma$/\Hb as a function of \FHb. \textbf{Upper panels:}  early type galaxies, \textbf{middle panels:}  spiral galaxies, \textbf{lower panels:} irregular galaxies. The number that appears at the top left corner of each panel indicates the number of PNe plotted in the panel. The number that appears at the top right corner of each panel indicates the weighted standard deviation of the H$\gamma$/\Hb values divided by its theoretical value. The numbering of the references is the same as in Fig.~\ref{fig:quality1}.}
  \label{fig:quality3}
\end{figure*}

In Figures~\ref{fig:quality1}--\ref{fig:quality3} we present the following three plots for all the galaxies: i) the [\ion{N}{ii}] $\lambda$6584/$\lambda$6548 ratio as a function of [N~{\sc ii}]$\lambda$6584/H$\beta\times$\FHb, ii) the [O~{\sc iii}] $\lambda$5007/$\lambda$4949 ratio as a function of [O~{\sc iii}] $\lambda$5007/H$\beta\times$\FHb, and iii) the H$\gamma$/\Hb ratio as a function of \FHb. These plots allow us to have an idea of the quality of the observational data. The number that appears at the top left corner of each panel indicates the number of PNe plotted in each panel. The upper panels correspond to early-type galaxies (M\,32, NGC\,4697, NGC\,205, and NGC\,147), the middle panels to spiral galaxies (M\,81, M\,31, M\,33, and NGC\,300, and NGC\,3109), and the lower panel to the irregular galaxies (LMC, SMC, NGC\,4449, and NGC\,6822). We have skipped plotting PNe with upper li\-mits in the line intensities as they are not useful for a quality check. In each of these three figures we also show (upper right corner) the weighted standard deviation of the values of [\ion{O}{iii}] $\lambda$4959/$\lambda$5007, [\ion{N}{ii}] $\lambda$6548/$\lambda$6584, and H$\gamma$/\Hb divided by the theoretical value. This quantity provide a more quantitative comparison between galaxies.

We have computed the error bars in the following plots through analytical error propagation. We have only considered the available uncertainties in the quantities involved in the calculations (i.e. basically those related to the uncertainties in line fluxes).

To distinguish between the various references we have used a color- and symbol-code, of course better seen when zooming on the figures. The color- and symbol-code are defined at the top right corner of each figure. The number scheme for each galaxy is as follows:

\begin{itemize}
    \item M\,31: (1) \citet[][B13]{balick13}, (2) \citet[][C15]{corradi15}, (3) \citet[][F15]{fang15}, (4) \citet[][F18]{fang18}, (5) \citet[][J99]{jacoby99}, (6) \citet[][K12]{kwitter12}, (7) \citet[][R99]{richer99}.
    \item M\,33: (1) \citet[][M09]{magrini09}, (2) \citet[][B10]{bresolin10}.
    \item NGC\,3109: (1) \citet[][FD17]{floresduran17}, (2) \citet[][P07]{pena07a}.
    \item LMC: (1) \citet[][LD06]{leisy06}, (2) \citet[][T03]{tsamis03}.
    \item SMC: (1) \citet[][LD06]{leisy06}, (2) \citet[][S10]{shaw10}, (3) \citet[][T03]{tsamis03}.
    \item NGC\,6822: (1) \citet[][GR16]{garciarojas16}, (2) \citet[][HM09]{hdezmtnez09}, (3) \citet[][R07]{richer07}.
\end{itemize}   

Departures of the [\ion{O}{iii}] $\lambda$4959/$\lambda$5007  ratios  from the theoretical value  that strongly exceed the error bars indicate that the estimated error bars are too small. This concerns especially some objects in M\,33 from the observations of \citet{magrini09}.

The error bars on the [\ion{N}{ii}] $\lambda$6548/$\lambda$6584 ratios are larger than those on the [\ion{O}{iii}] $\lambda$4959/$\lambda$5007  ratios because these lines are in general weaker than the \oiii\ lines and because of  the difficulty of deconvolving [\ion{N}{ii}] $\lambda$6548 from H$\alpha$, especially for high excitation objects observed with low spectral resolution. 
We note that large error bars on the [\ion{O}{iii}] $\lambda$4959/$\lambda$5007  ratios at low fluxes, especially in M\,31, M\,33, LMC, and SMC indicate that the estimations of the error bars in this case must have been too conservative, since most of the nominal values of this ratio fall close to the theoretical value.

Another way to appreciate differences between galaxies is to look at the values of the weighted standard deviation of these ratios divided by their theoretical values. This quantity is significantly higher in the case of [\ion{N}{ii}] $\lambda$6548/$\lambda$6584 than in that of [\ion{O}{iii}] $\lambda$4959/$\lambda$5007, as expected. The two exceptions are NGC~205 and NGC~147 with a similar value in both cases. The worst cases are M\,31, M\,33, LMC, NGC\,3109, NGC\,4449, and NGC\,6822 in Figure~\ref{fig:quality1} and M\,33, M\,81, and NGC\,3109 in Figure~\ref{fig:quality2}. Note that many of the galaxies have a few outliers in Fig.\ref{fig:quality1}. They should be considered with caution in the forthcoming papers.

Note that data from \citet{bresolin10} for M\,33 PNe are not reported in the plots of  [\ion{O}{iii}] $\lambda$4959/$\lambda$5007 and  [\ion{N}{ii}] $\lambda$6548/$\lambda$6584 because the weakest lines of these doublets are not listed in the original paper, but as can be judged from the spectra shown in that paper, the data are of excellent quality.

For the H$\gamma$/\Hb ratio, departures from the theoretical value exceeding the error bars may indicate an underestimation of the error bar in the H$\gamma$ emission line intensity, a problem in the spectral calibration over a wavelength range of about 500 \AA, or the fact that the determination of the stellar continuum in the region of H$\gamma$ does not take into account  stellar absorption with sufficient accuracy; however, this possibility is unlikely in most of the objects in our sample and should only be considered in galaxies with important old stellar populations or in the bulge of spiral galaxies like M\,31. The H$\gamma$/\Hb line ratio reported in the figure is in agreement with the theoretically expected value for some of the galaxies: M\,31, NGC\,300, NGC\,3109, LMC, SMC, either because the reddening was obtained using these lines precisely or because the observers paid special attention to it. But it is very far from it in  NGC\,205, M\,32, M\,81, and many objects in M\,33. The first five galaxies have a value of the weighted standard deviation of H$\gamma$/\Hb divided by its theoretical value $\lesssim0.01$ while this value is $\gtrsim0.03$ for NGC\,205, M\,32, M\,81, and M\,33. Whatever the cause, an incorrect H$\gamma$/\Hb ratio implies an incorrect [\ion{O}{iii}] $\lambda$4363/$\lambda$5007 ratio by about the same amount, which will affect the determination of the electron temperature.


\section{The real nature of the nebulae in our samples}
\label{nature}

In this section we examine whether the studied objects, which are considered as PNe in the original papers\footnote{For NGC\,3109, we have removed from our tables PN7, following the arguments presented by \citet{pena07a}. Similarly, for NGC\,300, we included object 74 and discarded object 39 following the strong arguments presented by \citet{stasinska13}.}, could be something else  such as giant \hii regions in late-type galaxies, or supernova remnants (SNRs) in all types. 

We also try to get a hint on the behaviour of the density: are the densities derived by \ariv\ and \sii\ systematically different? In case only one of the two is available, can it be used as representative of the whole nebula for nebular mass estimations and for abundance ratios calculations? 

\begin{figure*}
  \flushleft
  \includegraphics[width=1.0\textwidth, trim= 20 0 40 0cm]{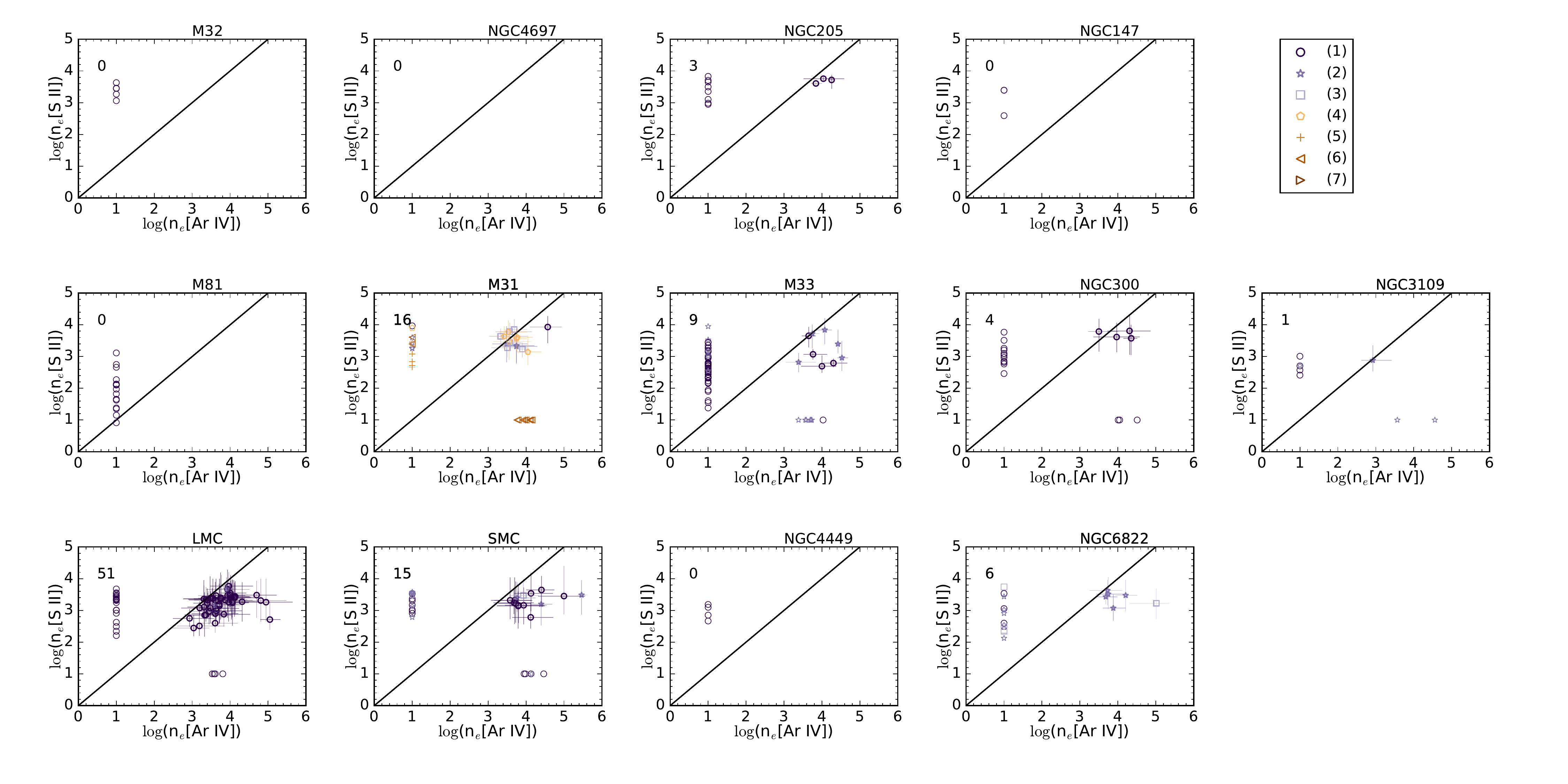}
  \caption{Values of \denssii\ $\lambda$6731/$\lambda$6717 as a function of \densariv\ $\lambda$4740/$\lambda$4711. \textbf{Upper panels:}  early type galaxies, \textbf{middle panels:}  spiral galaxies, \textbf{lower panels:} irregular galaxies. The number that appears at the top left corner of each panel indicates the number of PNe plotted in each panel. When densities for a given line-ratio were not available, they were systematically put to 10. The numbering of the references is the same as in Fig.~\ref{fig:quality1}.}
  \label{fig:densities1}
\end{figure*}

Figure~\ref{fig:densities1} shows the values of \denssii\ $\lambda$6731/$\lambda$6717 as a function of \densariv\ $\lambda$4740/$\lambda$4711. Both values were derived with {\sc PyNeb} v.1.1.10 \citep{Luridiana15}. The number that appears at the top left corner of each panel indicates the number of objects plotted in the panel (upper limits or objects that do not have a value of the two parameters plotted are not counted, but these objects are represented in the plots at a value of density equal to 10 cm$^3$).
As in the previous figures, the upper panels correspond to early-type galaxies, the middle panels to spiral galaxies, and the lower panels to the irregular galaxies. The color- and symbol-code for the references is the same as in previous figures. The atomic data we have adopted to compute the densities, as well as the correction for the \ariv\ $\lambda$4711 line are shown in Table~\ref{tab:atomic}.

\begin{table}
\caption{Atomic data set used for collisionally excited lines and recombination lines.}
\label{tab:atomic}
\begin{adjustbox}{width=\columnwidth}
\begin{tabular}{ccc}
\hline
\multicolumn{3}{c}{Collisionally excited lines}\\ 
\hline
\multicolumn{1}{l}{Ion} & \multicolumn{1}{c}{Transition Probabilities} &
\multicolumn{1}{c}{Collision Strengths} \\
\hline
O$^{2+}$  &  \citet{Froesefischer04}, & \citet{Storey14}\\
   &  \citet{Storey00} & \\
S$^{+}$   &  \citet{Mendoza82b} & \citet{Tayal10}\\
Ar$^{3+}$ &   \citet{Mendoza82b}  & \citet{Ramsbottom97}\\
\hline
\multicolumn{3}{c}{Recombination lines} \\
\hline
\multicolumn{1}{l}{Ion} & \multicolumn{2}{c}{Effective recombination Coefficients} \\
\hline
H$^{+}$ & \multicolumn{2}{c}{\citet{Storey95}} \\
He$^{+}$ & \multicolumn{2}{c}{\citet{Porter12,Porter13}}\\
\hline
\end{tabular}
\end{adjustbox}
\end{table}

In some cases there are two determinations for the densities, through \sii\ and \ariv. In most cases, there is only one determination, and in some cases no determination at all. Most of our objects have densities between 1\,000 and 10\,000 cm$^{-3}$. When both \denssii\ and \densariv\ are available, the density derived from \ariv\ tends to be larger, typically by a factor of $\sim$10 but sometimes by up to a factor of $\sim$200. 
We have checked the effect of self-absorption in the two He~{\sc i} lines we used to compute the theoretical ratio to correct the \ariv\ $\lambda$4711 line, He~{\sc i} $\lambda$4713 and $\lambda$5876, but this ratio would be affected in the opposite direction, i.e. if the effect is not considered in the correction, we would have lower densities.

The relation between \denssii\ and \densariv\ that we see in Fig. \ref{fig:densities1} could be due to the density structure of the objects, with the density decreasing outwards. But this seems in contradiction with the recent finding by \citet{rodriguez20} that in Galactic PNe those two densities are similar. However, one must recall that in the case of Galactic PNe, most of the times the observing aperture covers only a small portion of the object -- generally chosen to be the brightest --, so the  density measurements in Galactic and extragalactic PNe do not exactly mean the same. This subject needs to be investigated further, as the value adopted for the electron densities in the calculation of nebular masses or of some abundance ratios is important. 

\begin{figure*}
  \flushleft
  \includegraphics[width=1.0\textwidth, trim= 20 0 40 0cm]{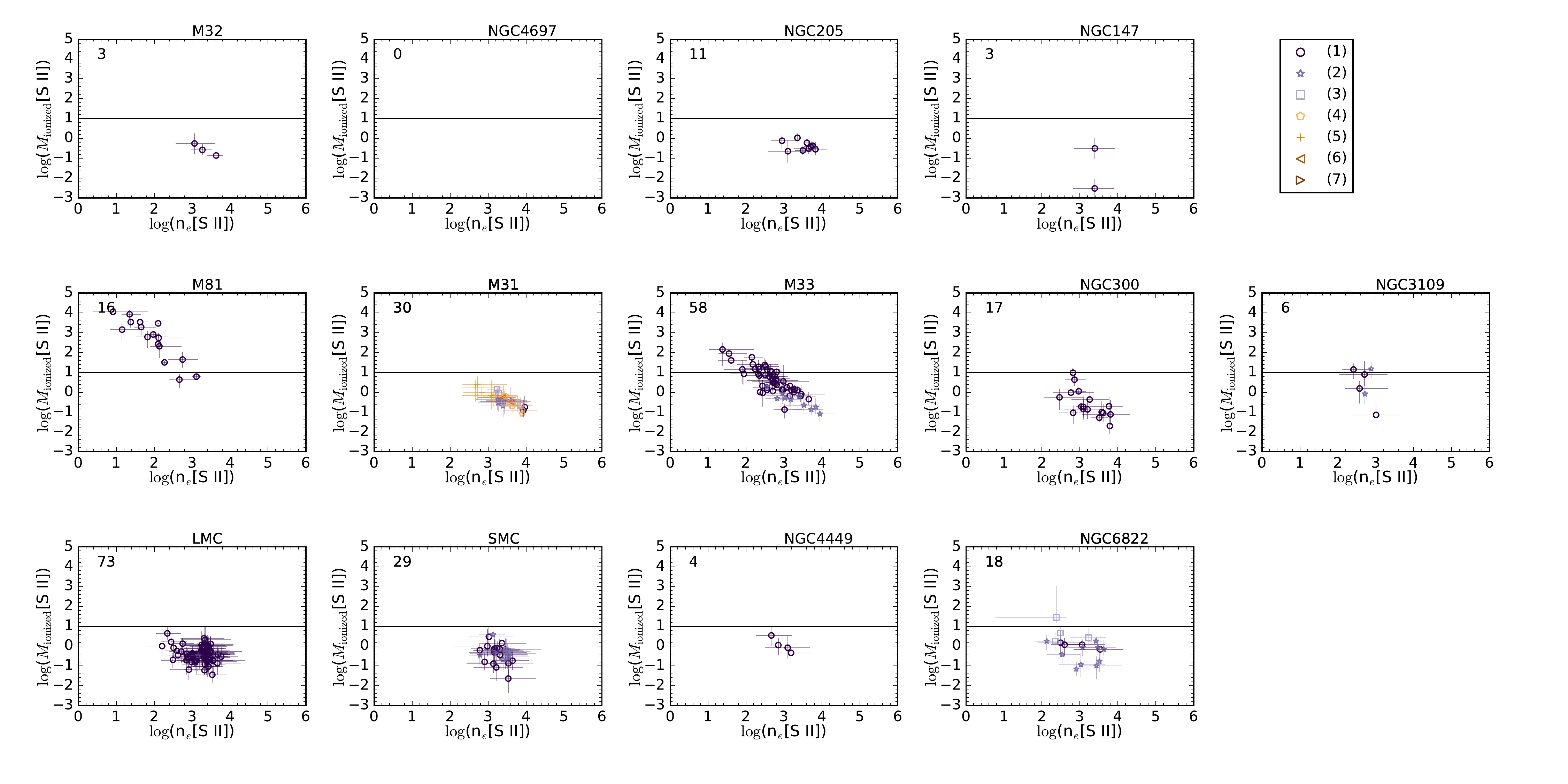}
  \caption{Values of the ionized mass as a function of the electron density derived both from \sii\ lines. \textbf{Upper panels:}  early type galaxies, \textbf{middle panels:}  spiral galaxies, \textbf{lower panels:} irregular galaxies. The number that appears at the top left corner of each panel indicates the number of PNe plotted in each panel. The numbering of the references is the same as in Fig.~\ref{fig:quality1}.}
  \label{fig:densities2}
\end{figure*}

\begin{figure*}
  \flushleft
  \includegraphics[width=1.0\textwidth, trim= 20 0 40 0cm]{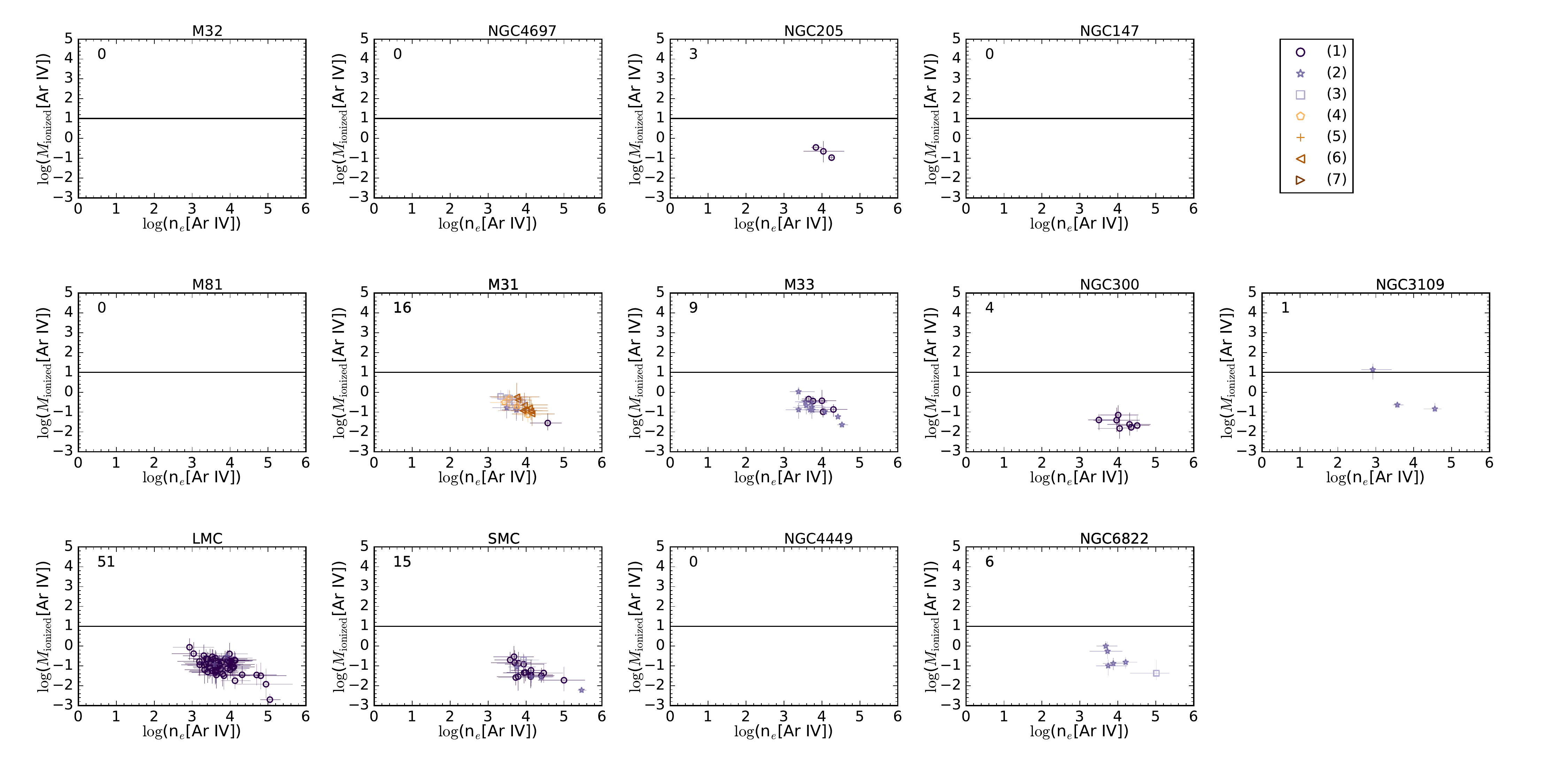}
  \caption{Values of the ionized mass as a function of the electron density derived both from \ariv\ lines. \textbf{Upper panels:}  early type galaxies, \textbf{middle panels:}  spiral galaxies, \textbf{lower panels:} irregular galaxies. The number that appears at the top left corner of each panel indicates the number of PNe plotted in each panel. The numbering of the references is the same as in Fig.~\ref{fig:quality1}.}
  \label{fig:densities3}
\end{figure*}

Figure~\ref{fig:densities2} and \ref{fig:densities3} show the values of the ionized mass as a function of the electron density derived from \sii\ and \ariv\ lines, respectively. The ionized masses have been derived from the equation given by \citet{stasinska13}: 
\begin{equation}
\label{mass}
M_{\rm ionized} = 37.5   L(\Hb) / n_e  ,
\end{equation}
 where  $L(\Hb)$ and $M_{\rm ionized}$ are in solar units, and $n_e$ is the mass-weighted average electron density in cm$^{-3}$. In practice, we use the \sii\ or \ariv\ ratios to estimate this density.

Since we do not know which of \denssii\ or \densariv\ better represents the bulk density of the ionized gas, we estimated the masses of the ionized gasses using both densities in turn.

In PNe, the ionized masses cannot be larger than a few solar masses at the very most, since PNe arise from intermediate-mass stars. The determination of the ionized mass is straightforward if the density is known, and does not depend on the geometry. In Eq. \ref{mass} the density $n_e$ represents some average density of the object, which may not be identical with the one measured by \sii\ or \ariv\ line diagnostics. The densities derived from \sii\ and \ariv\ may actually be higher than the average density of the whole nebula, so the real $M_{\rm ionized}$ may be higher than estimated.

\begin{figure*}
\flushleft
\includegraphics[width=\textwidth, trim= 20 0 40 0cm]{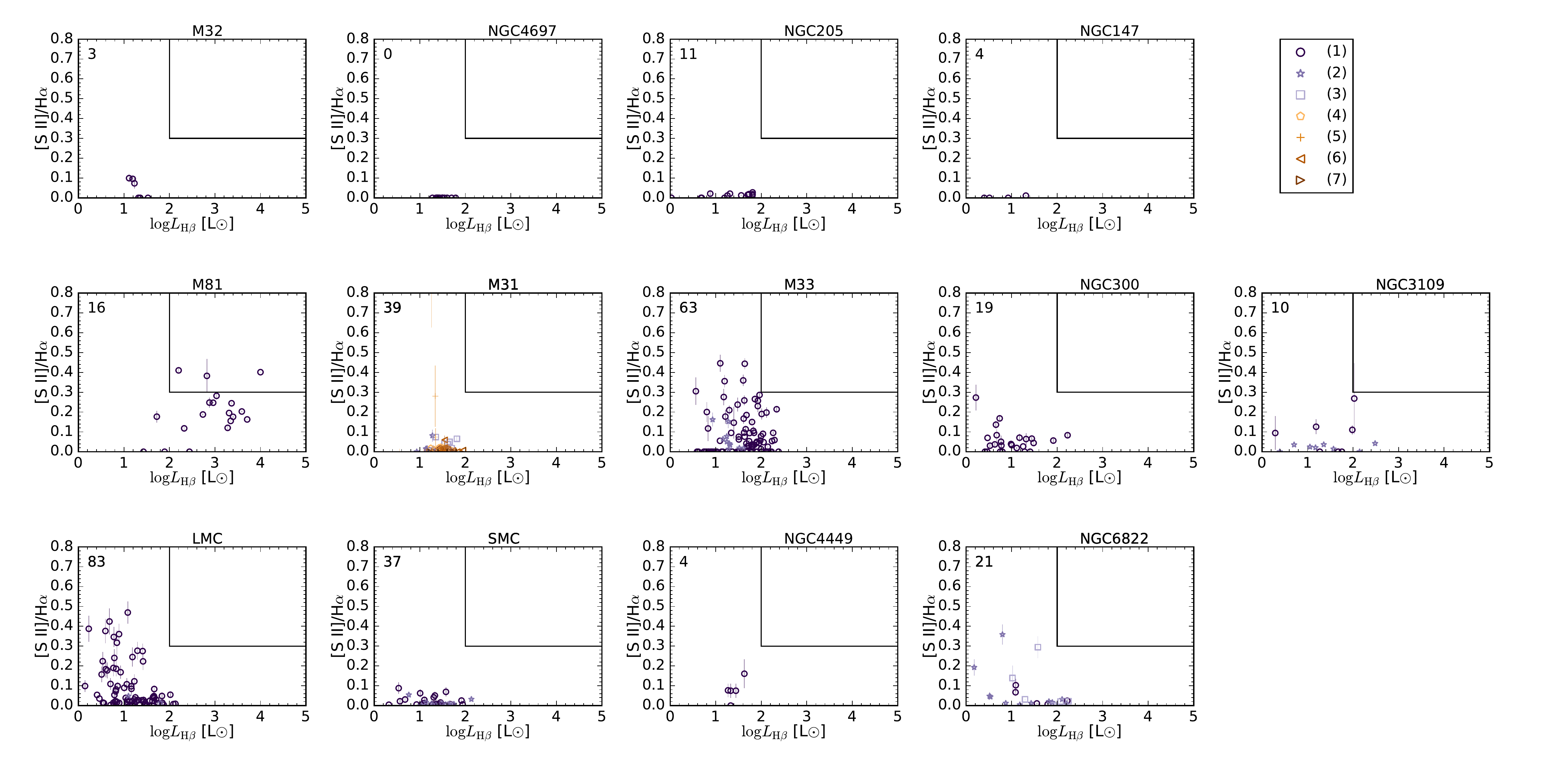}
\caption{Values of \sii/\Ha as a function of \LHb. \textbf{Upper panels:}  early type galaxies, \textbf{middle panels:}  spiral galaxies, \textbf{lower panels:} irregular galaxies. The number that appears at the top left corner of each panel indicates the number of PNe plotted in each panel. The upper right box defines the region where SNRs are expected: \sii/\Ha > 0.3 and $\log$(\LHb) > 2. The numbering of the references is the same as in Fig.~\ref{fig:quality1}.}
\label{fig:diagnostic1}
\end{figure*}

But objects for which the derived nebular mass is larger than, say, 10 \msun (this value is marked with a line in the corresponding plots), definitely cannot be PNe. There are some of such objects in NGC\,6822 (PN13 and PNS16) and M\,33 (PN29, PN30, PN32, PN47, PN48, PN51, PN52, PN59, PN67, PN68, PN90, PN92, and PN94), and it is the case of the majority of objects in M\,81 for which the densities were determined (see Figure~\ref{fig:densities1}). The limit of 10 we considered is very conservative. Note also that for those objects, the \Hb luminosities are higher by one or 2 orders of magnitude than for the rest of our sample (see Figure~\ref{fig:diagnostic1} to be described later). This indicates that these objects are not PNe and are likely compact \hii regions. 

For NGC\,300, \citet{roth18} classified 5 out of the 18 objects in common with \citet{pena12} as compact \hii region candidates and not PNe. They did this on the basis of the high \sii/\Ha  ratios observed in these objects, a criterion  wrongly attributed to \citet{stasinska13}. Our determinations of ionized masses suggests that all the objects that \citet{stasinska13} considered as are indeed PNe. 

In Figures~\ref{fig:densities2} and \ref{fig:densities3}, we see that, in some galaxies, there is an anticorrelation between the ionized masses and the electron densities. This clearly happens for M\,31, M\,33, M\,81, and NGC\,300. For a given number of hydrogen ionizing photons, $Q_H$, the mass of gas that can be ionized is proportional to $Q_H$/$n_e$. The behaviour that we see between $M_{\rm ionized}$ suggests that most objects are ionization bounded. This, of course, cannot be considered as fully certain, since we know that the values of $Q_H$ vary from object to object among PNe, and a deeper analysis is required. However, we note that the objects in the LMC and SMC do not show this trend at all (this is more clearly seen in the plots involving \sii\ lines), which suggests that many of them are density-bounded. It is remarkable that such a conclusion has already been reached by Barlow (1987) -- though not on the same sample and using different arguments.

Can some of the objects considered in this study be actually SNRs and not PNe? This concern was expressed by \citet{Davis2018}. From a diagram plotting \oiii/(\Ha + \nii) versus the absolute magnitude in \Oiii, these authors concluded that  SNRs were not likely to contaminate the upper part of the PN luminosity function in  M\,31 and M\,33. 

Figure~\ref{fig:diagnostic1} shows the values of \sii/\Ha as a function of \LHb\ for our objects. From a census of optical data on extragalactic SNRs in some of the galaxies of our sample \citep{Long2010, Leonidaki13, Lee14}, SNRs always have \sii/\Ha larger than 0.3 (see Fig.~\ref{fig:snr} in Appendix~\ref{app2}). This implies that objects with \sii/\Ha smaller than 0.3 in our sample cannot be SNRs. Some of our objects do have \sii/\Ha larger than 0.3, but this is not considered incompatible with a PN nature (see Fig.7 of \citealt{sabin13}). On the other hand, SNRs in external galaxies have higher luminosities than PNe. As shown in Fig. \ref{fig:snr}, 5 SNRs out of 560 have $\log$\LHb smaller than 2 (and these values are not corrected for extinction, contrary to the values of \LHb\ for our samples). The upper right box in figure~\ref{fig:diagnostic1} defines the region where SNRs are expected: \sii/\Ha > 0.3 and $\log$\LHb > 2. As this figure shows, in our samples, we do not see any object in this region (except a few objects in M\,81 for which we already know they are not PNe).

We note that the distinction between PNe and other types of nebulae is not necessarily clear-cut from an analysis such as the one we performed here. More information can come from a study of the chemical composition, to be presented in a future paper.


\section{PN evolution}
\label{plots}

\subsection{Luminosities, densities and extinction}
\label{lumdensext}

\begin{figure*}
\flushleft
\includegraphics[width=\textwidth, trim= 20 0 40 0cm]{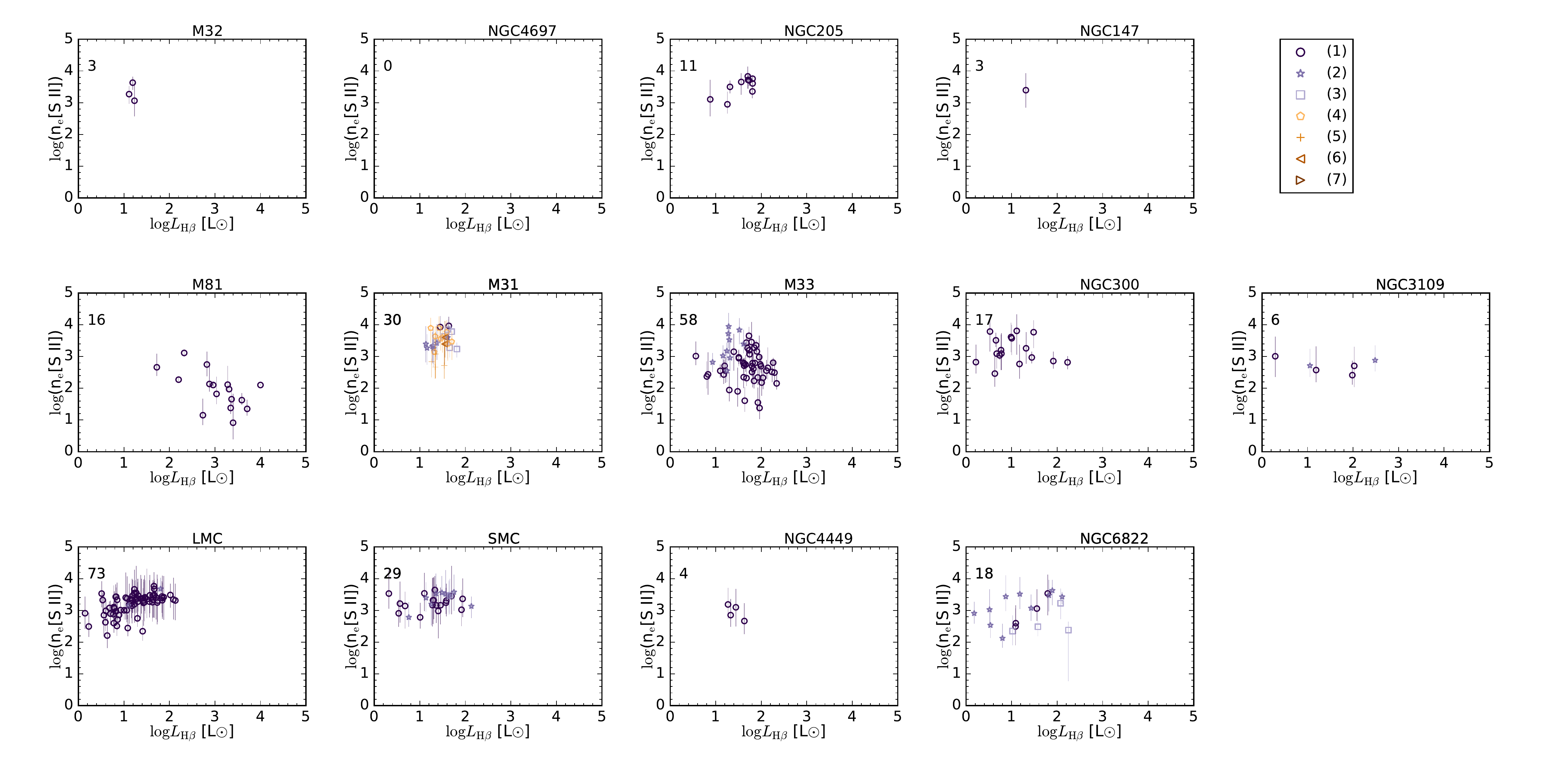}
\caption{Values of \denssii\ as a function of \LHb. \textbf{Upper panels:}  early type galaxies, \textbf{middle panels:}  spiral galaxies, \textbf{lower panels:} irregular galaxies. The number that appears at the top left corner of each panel indicates the number of PNe plotted in each panel. The numbering of the references is the same as in Fig.~\ref{fig:quality1}.}
  \label{fig:lumdensext1}
\end{figure*}

\begin{figure*}
\flushleft
\includegraphics[width=\textwidth, trim= 20 0 40 0cm]{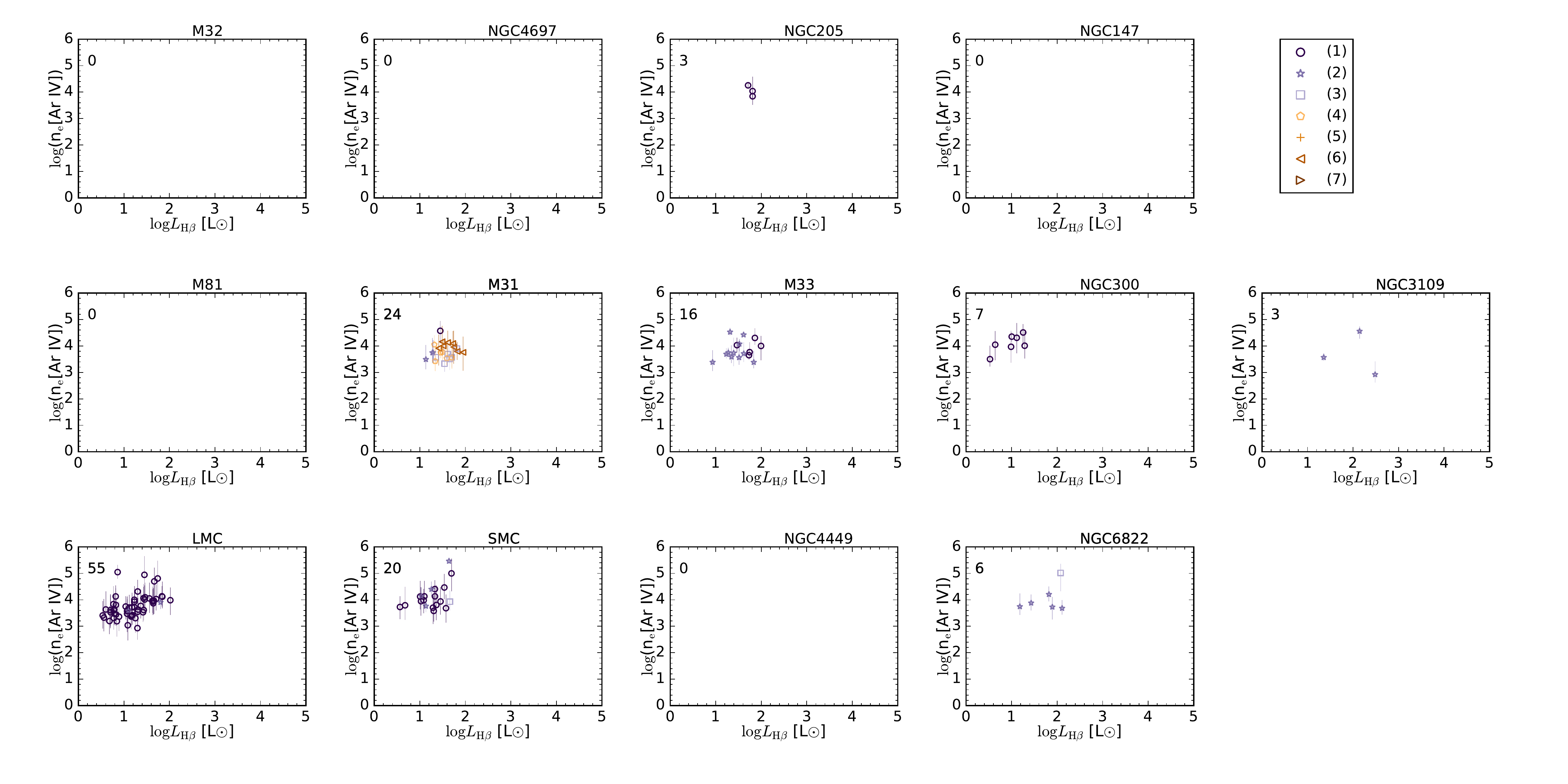}
\caption{Values of \densariv\ as a function of \LHb. \textbf{Upper panels:}  early type galaxies, \textbf{middle panels:}  spiral galaxies, \textbf{lower panels:} irregular galaxies. The number that appears at the top left corner of each panel indicates the number of PNe plotted in each panel. The numbering of the references is the same as in Fig.~\ref{fig:quality1}.}
  \label{fig:lumdensext2}
\end{figure*}

Figures~\ref{fig:lumdensext1} and \ref{fig:lumdensext2} show the densities \denssii\ and \densariv\ as a function of \LHb. If all the nebulae were to have a si\-mi\-lar total mass (including the non-ionized part) one would expect the objects with the lowest densities to correspond to the oldest nebulae. Such an argument is often mentioned in studies of PNe in the Milky Way \citep{Osterbrock2006}. It comes naturally from the idea that PNe are expanding. If the brightest PNe were also among the youngest, as is sometimes thought \citep[e.g.,][]{richer06} one would expect the densities to be correlated with \LHb. 
We see this clearly for the LMC and the SMC, but not for the rest of the galaxies.
In general there is no correlation, and there even seems to be a slight anticorrelation in M\,33. 
As a matter of fact, \LHb\ is not a decreasing function of time. For a given object, as long as it is ionization-bounded, \LHb\ is proportional to the rate of variation of the total number of ionizing photons, which, in the first stages of PN evolution first increases an then decreases. As the object becomes density-bounded, \LHb\ becomes directly proportional to the density. When considering a collection of objects with different central star masses, as is the case here, the observed plots result from a mixture of time scales, since more massive central stars evolve more rapidly. In addition, the mass-loss rates at the tip of the AGB as well as the expansion velocities also play a role.
A proper interpretation of the behaviour of our samples in the $n_e$--\LHb\ plane therefore requires a modelling approach such as in  \citealt{stasinska98} (a study not taking into account dynamical effects), or, even better,  \citealt{schonberner07} who present a fully hydrodynamical modelling. This will be attempted in forthcoming publications.

\begin{figure*}
\flushleft
\includegraphics[width=\textwidth, trim= 20 0 40 0cm]{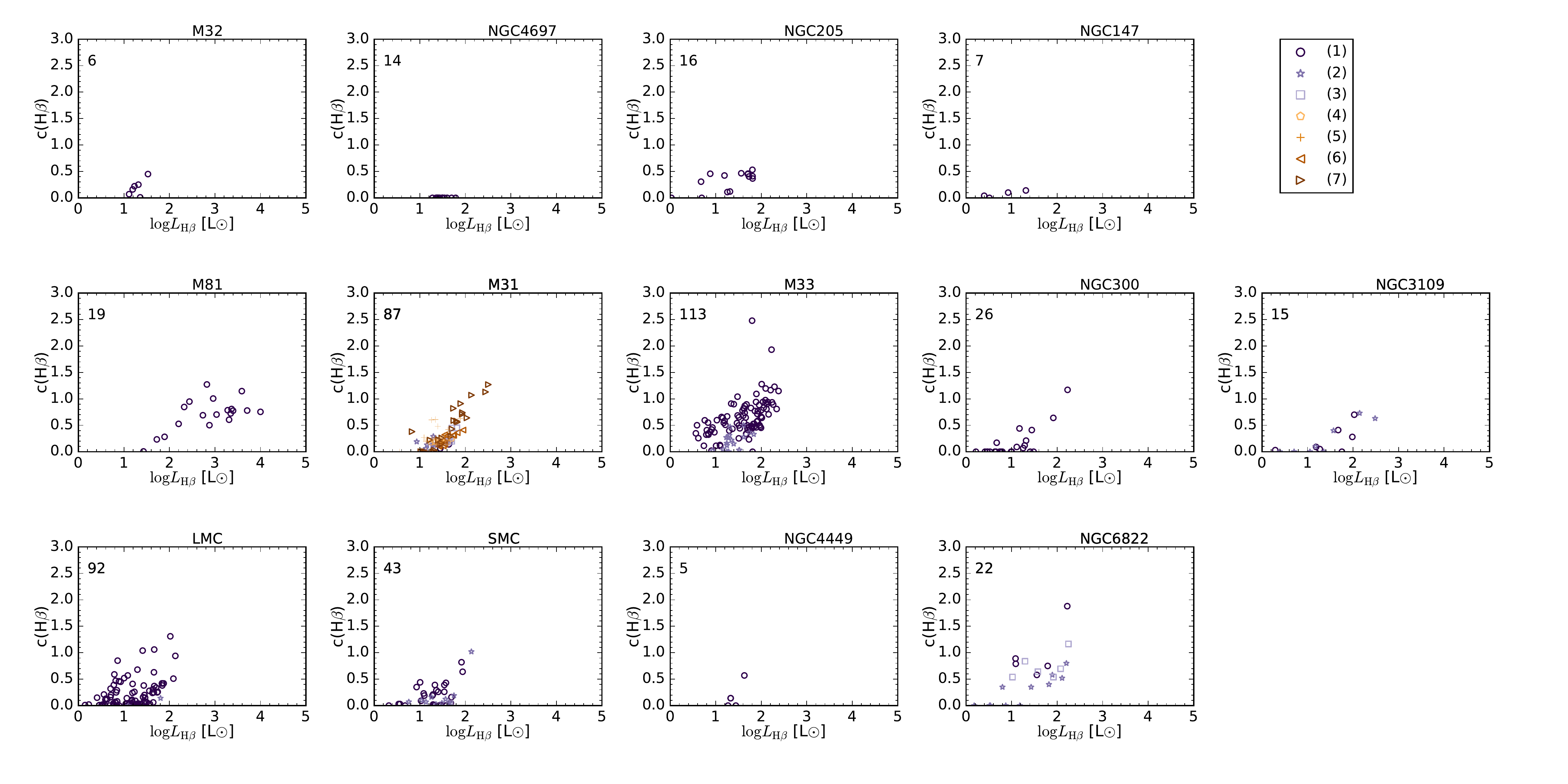}
\caption{Values of \cHb\ as a function of \LHb. \textbf{Upper panels:}  early type galaxies, \textbf{middle panels:}  spiral galaxies, \textbf{lower panels:} irregular galaxies. The number that appears at the top left corner of each panel indicates the number of PNe plotted in each panel. The numbering of the references is the same as in Fig.~\ref{fig:quality1}.}
  \label{fig:lumdensext3}
\end{figure*}

Figure~\ref{fig:lumdensext3} show the value of the extinction \cHb\ as a function of \LHb. In M\,31, M\,33, NGC\,300, LMC and NGC\,6822\footnote{We exclude M\,81 from this consideration, since we have already shown that most -- if not all -- the objects of the sample are not PNe.} we see a clear increase of \cHb\  with increasing \LHb. It has already been argued by \citet{ciardullo99}, on the basis of observations in M\,31,  that PNe with progenitors of higher stellar mass have larger extinction. We postpone the question of progenitor masses to a future paper, but we can state that our empirical finding -- which appears to be very general -- is very important in the context of PN luminosity function studies and their application to the derivation of galaxy distances. 

\subsection{Evolution of line ratios}
\label{lineratios}

\begin{figure*} 
\flushleft
\includegraphics[width=\textwidth, trim= 20 0 40 0cm]{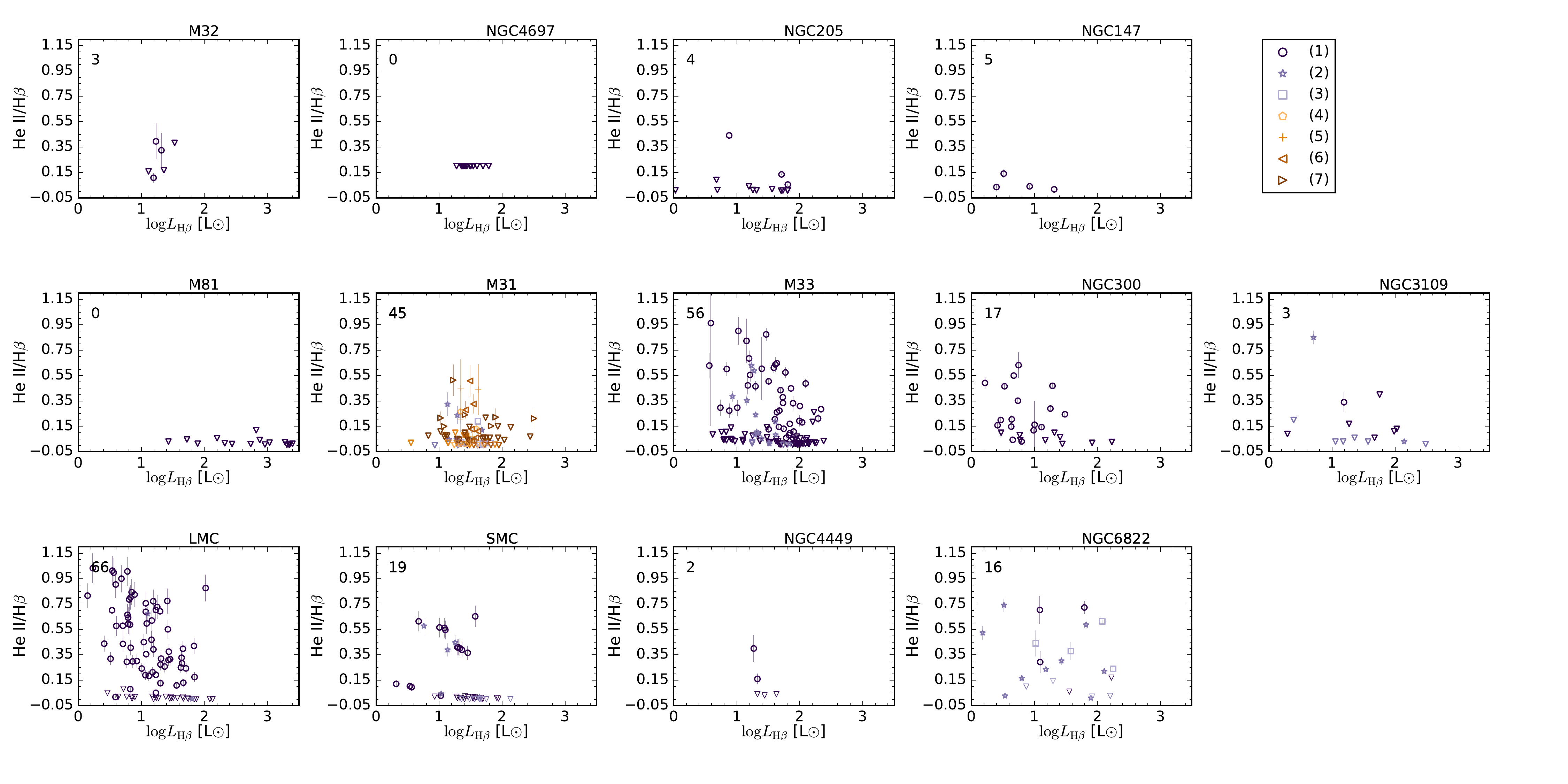} \caption{Values of \Heii/\Hb as a function of $\log$\LHb. \textbf{Upper panels:}  early type galaxies, \textbf{middle panels:}  spiral galaxies, \textbf{lower panels:} irregular galaxies. The number that appears at the top left corner of each panel indicates the number of PNe plotted in each panel. The numbering of the references is the same as in Fig.~\ref{fig:quality1}.}
  \label{fig:evolutionary1}
\end{figure*}

\begin{figure*}
\flushleft
\includegraphics[width=\textwidth, trim= 20 0 40 0cm]{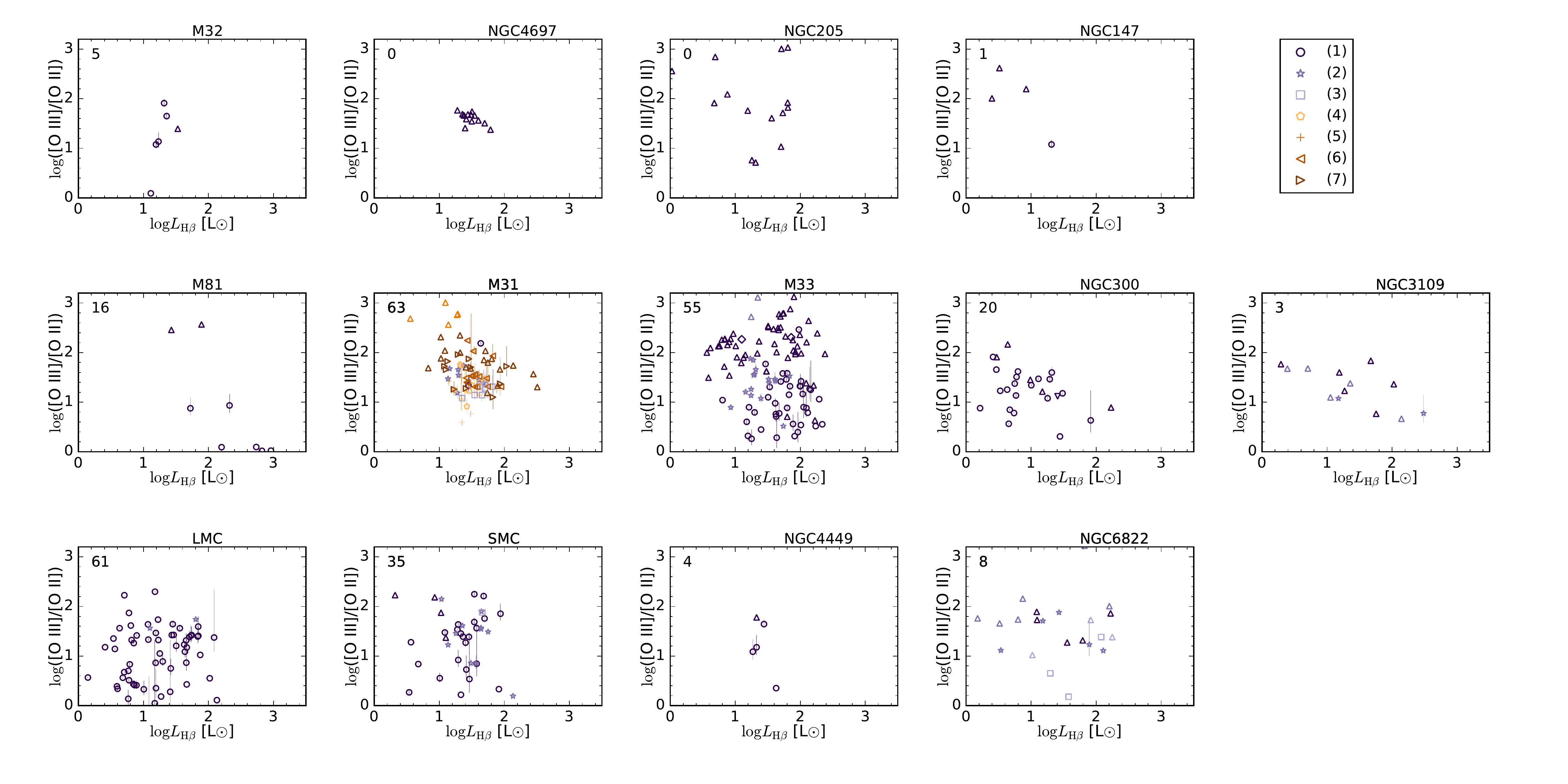}
\caption{Values of $\log$\Oiii/\Oii\ as a function of  $\log$\LHb. \textbf{Upper panels:}  early type galaxies, \textbf{middle panels:}  spiral galaxies, \textbf{lower panels:} irregular galaxies. The number that appears at the top left corner of each panel indicates the number of PNe plotted in each panel. The numbering of the references is the same as in Fig.~\ref{fig:quality1}.}
  \label{fig:evolutionary2}
\end{figure*}

Figures~\ref{fig:evolutionary1} and \ref{fig:evolutionary2} show the variations of \Heii/\Hb and  \Oiii/\Oii\ as a function of \LHb. These ratios do not depend on abundances (at least directly). The panels have been arranged following the criteria in previous plots, i.e. by increasing order of the de Vaucouleurs index from RC3 (i.e., earliest types on the left-top panels, latest types on the right-down panels). Different sets of observations for the same galaxy are represented by different colours, as indicated before. The vertical segments represent the error bars. Upper limits are marked by downwards pointing triangles, lower limits by upwards pointing triangles. 

One striking aspect of Figure~\ref{fig:evolutionary1} is the behaviour of \Heii/\Hb in the different samples. This line ratio depends on the effective temperature of the star and on the ionized/neutral nebular mass ratio. It was already noted by \citet{stasinska98} that PNe in M\,32 and in the bulge of M\,31 do not contain objects with high values of \Heii/\Hb, whereas in the LMC there are many objects in which this ratio is much larger that 0.6. This does not mean that the central stars of these objects are more massive and reach higher temperatures, as one might think. As a matter of fact in an ionization bounded nebula, the \Heii/\Hb ratio cannot exceed, say, 0.9, and for stellar temperatures that are lower than 200,000K \citep[which is what is expected for most of PNs according to stellar evolution models, see][]{blocker95, millerbertolami16} it is even smaller than 0.6 (see~Figure~\ref{fig:models} in the Appendix \ref{app1}). This means that in the LMC and in M\,33, and possibly also SMC and NGC\,6822, many PNe are density bounded. Such a finding strengthens the diagnostic given in Sect. \ref{nature}, based on completely different arguments. The reason of this behavior will be investigated in a future paper. Note that (except in NGC\,6822) the highest values of \Heii/\Hb are not found for the PNe with highest \Hb luminosities. Note also that for M\,81, none of the objects has a measured value of \Heii, all the upper limits indicate \Heii/\Hb smaller than 0.12. This led \citet{stanghellini10} to interpret this fact as due to moderately low central star temperature (less than $\sim$100,000 K). A  more convincing explanation is that these objects are not PNe but compact \hii regions, as argued in the preceding section.

Figure~\ref{fig:evolutionary2} can be seen as the excitation as a function of \Hb luminosity. Unfortunately, in many objects of our samples, only a lower limit is available, due to the weakness of \oii\ lines. The observations, apart from showing a range of over three decades in \Oiii/\Oii\  do not present any clear trend in most of the panels and do not reveal any strongly different behaviour among the various galaxies. We can just note that the early type galaxies show consistently the smallest values of \Oiii/\Oii . Also, in the SMC we do not find such high ratios as in the LMC and M\,33, where  the PNe with the largest values (or lower limits) of \Oiii/\Oii\ are found. The \Oiii/\Oii\ intensity ratio depends on many parameters: the effective temperature of the star, the mean ionization parameter, and the ionized/neutral nebular mass ratio. This complex dependence cannot be deciphered without the help of appropriate models. In addition, we must recall that for several samples, a \Oiii/\Hb\ ratio larger than 3--4 is required for the object to be counted as a PN. 

\subsection{PN luminosities and the PN luminosity function}
\label{pnlf}

So far, in terms of luminosities, we only considered \LHb, the total luminosity in the \Hb\ line, corrected for extinction. The reason is that this is the physical parameter easiest to interpret \citep[see e.g.][]{dopita92, stanghellini95}. It depends on the number of ionizing photons emitted by the central star, on the nebular mass and its mean density. We could as well have used \LHa, which is about three times larger and depends on the same parameters. However, observational studies based on PN luminosities use the observed \oiii\ magnitudes instead, being \Oiii\ the brightest line in such objects. Actually,  the detection of extragalactic PNe mainly relies on \oiii. Many studies of the PN luminosity function (PNLF) concluded that the bright-end cutoff luminosity of
the PNLF is  invariant with respect to galaxy type and position in the galaxy, allowing the use of the PNLF as a tool to measure galaxy distances \citep[see][for a recent review]{ciardullo16}. As expressed in that paper, it is not yet clear why this technique works so well, because the observed \oiii\ luminosity, $L^{\rm obs}_{\oiii}$, also depends on metallicity, excitation and reddening.

\begin{figure*}
\flushleft
\includegraphics[width=\textwidth, trim= 20 0 40 0cm]{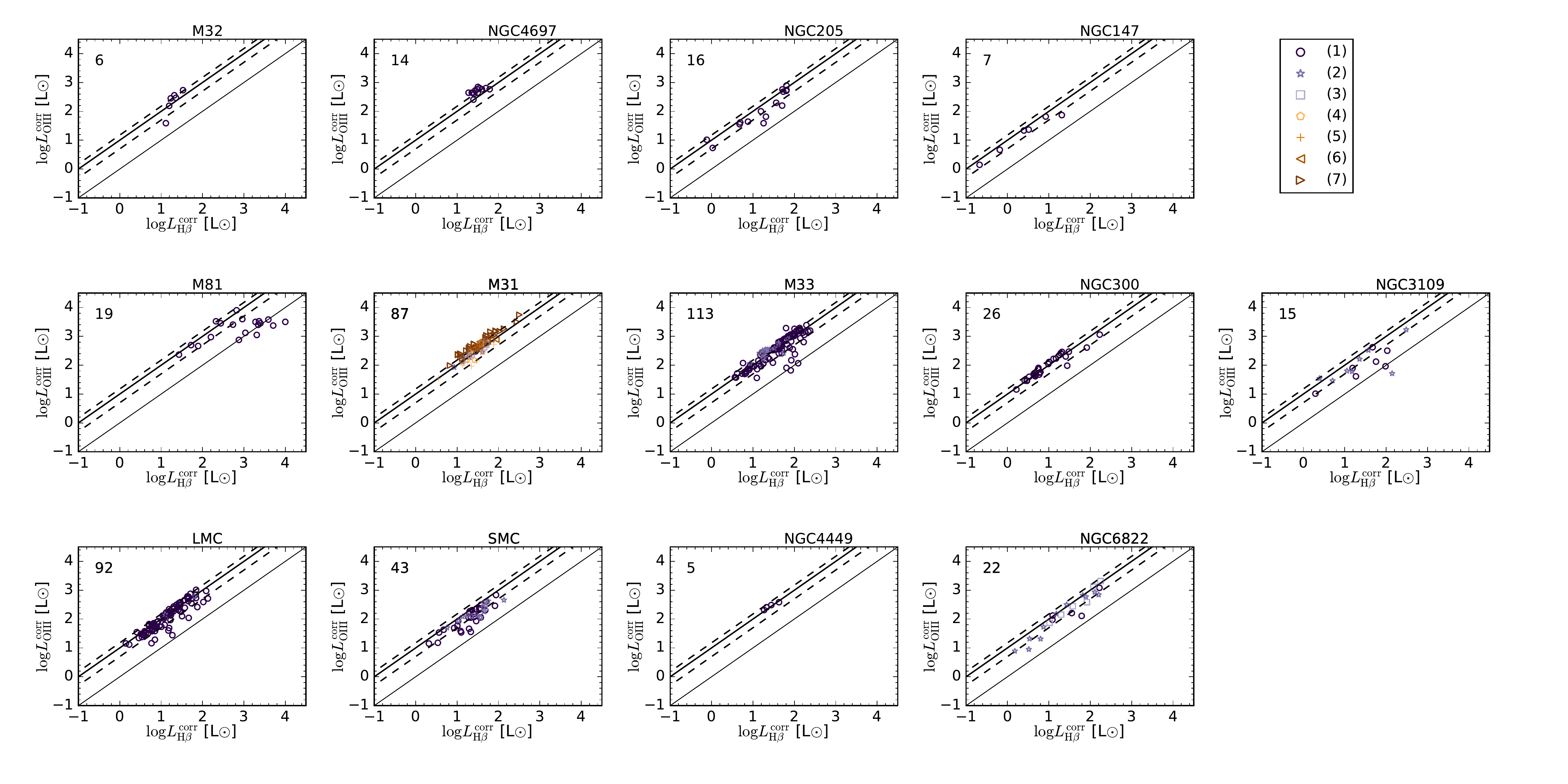}
\caption{Values of $\log L^{\rm corr}_{\oiii}$ as a function of $\log L^{\rm corr}_{\Hb}$. \textbf{Upper panels:}  early type galaxies, \textbf{middle panels:}  spiral galaxies, \textbf{lower panels:} irregular galaxies. The number that appears at the top left corner of each panel indicates the number of PNe plotted in each panel. The thin line represents where $L^{\rm corr}_{\oiii}$ = $L^{\rm corr}_{\Hb}$ whereas the thick solid line and the dashed lines are obtained from the mean and the standard deviation of $L^{\rm corr}_{\oiii}$/$L^{\rm corr}_{\Hb}$. The numbering of the references is the same as in Fig.~\ref{fig:quality1}.}
  \label{fig:evolutionary3}
\end{figure*}

\begin{figure*}
\flushleft
\includegraphics[width=\textwidth, trim= 20 0 40 0cm]{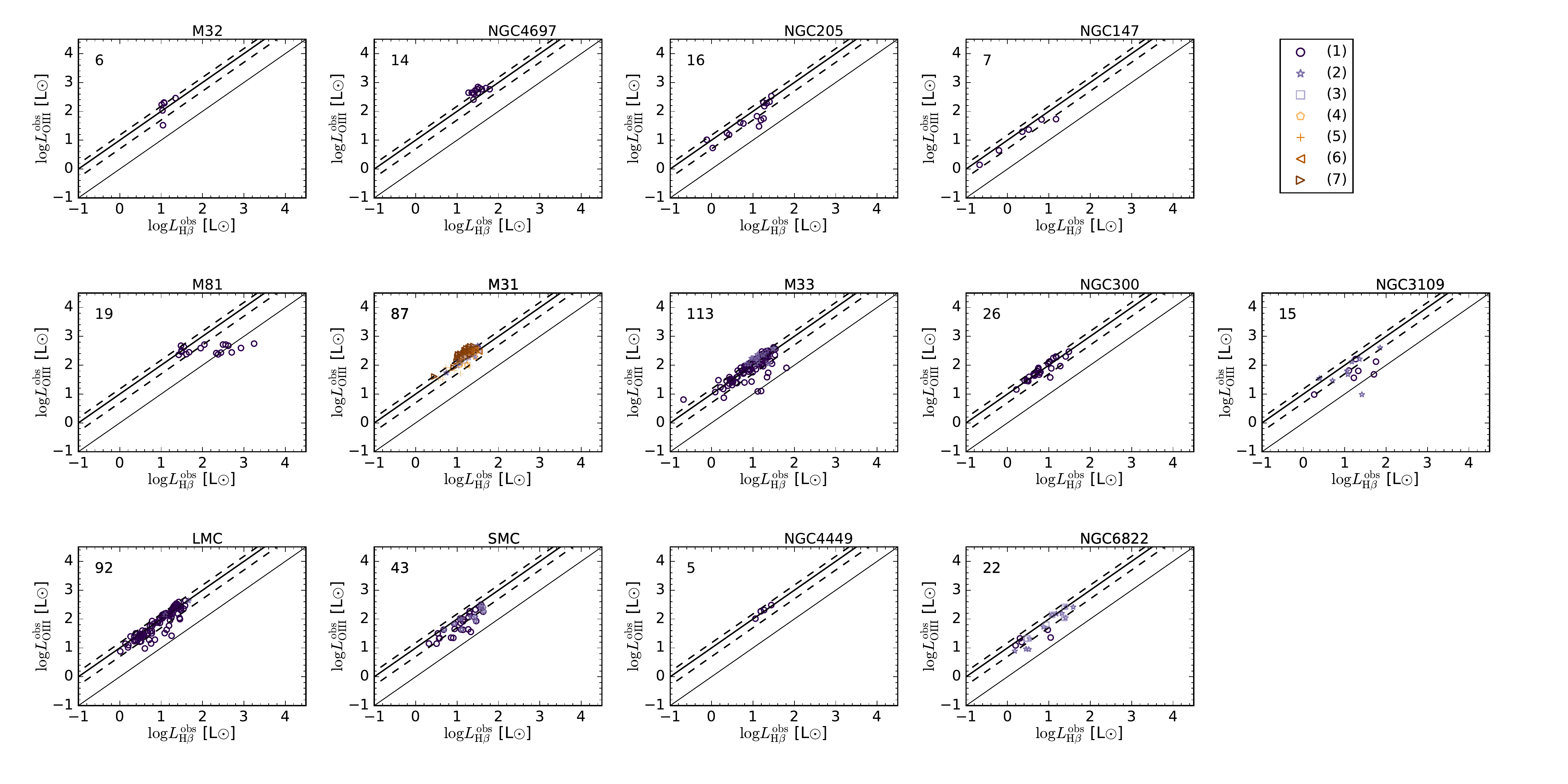}
\caption{Values of $\log L^{\rm obs}_{\oiii}$ as a function of $\log L^{\rm obs}_{\Hb}$. \textbf{Upper panels:}  early type galaxies, \textbf{middle panels:}  spiral galaxies, \textbf{lower panels:} irregular galaxies. The number that appears at the top left corner of each panel indicates the number of PNe plotted in each panel. The thin line represents where $L^{\rm obs}_{\oiii}$ = $L^{\rm obs}_{\Hb}$; the thick solid line and the dashed lines are obtained from the mean and the standard deviation of $L^{\rm obs}_{\oiii}$/$L^{\rm obs}_{\Hb}$. The numbering of the references is the same as in Fig.~\ref{fig:quality1}.}
  \label{fig:loiiilhbnotcorr}
\end{figure*}


In Figs. \ref{fig:evolutionary3} and \ref{fig:loiiilhbnotcorr} we show the relation between the logarithms of the  \oiii\ and \Hb\ luminosities, corrected and uncorrected for extinction, respectively. In both figures, the thin line represents the one-to-one relation. The thick solid line and the dashed lines are obtained from the mean and the standard deviation of $L_{\oiii}$/$L_{\Hb}$.

Figs. \ref{fig:evolutionary3} and \ref{fig:loiiilhbnotcorr} show that, for most PNe of our samples, the relation $L_{\oiii}$/$L_{\Hb}$ $= 10\pm5$ holds. Only about 14\% of the PNe have $L^{\rm corr}_{\oiii}$/$L^{\rm corr}_{\Hb}$ below 5. Those are objects of lower excitation than the majority and they can be found at any value of $L^{\rm corr}_{\Hb}$ in the diagrams, even at the highest ones  (e.g. in NGC\,3109 or the LMC). This, a priori, could affect the \oiii\ PNLF and its use for distances. As mentioned before, our samples are probably biased against such objects. In some galaxies (M\,32, NGC\,4697, M\,31), the majority of the PNe are slightly above the thick line, while in the SMC and NGC\,6822, they are slightly below. This is likely related to the metallicity, and will be studied in more detail in forthcoming papers. 

In Fig. \ref{fig:loiiilhbnotcorr}, the objects are displaced along the diagonal to the left with respect to their position in Fig. \ref{fig:evolutionary3} because the values of $f$($\lambda$) are very similar for \oiii\ and \Hb.   Since, as seen in Fig. \ref{fig:lumdensext3}, the extinction tends to  be larger for larger values of \LHb, the points in Fig. \ref{fig:loiiilhbnotcorr} are less spread along the diagonal direction than in Fig. \ref{fig:evolutionary3}. 
In Fig. \ref{fig:loiiilhbnotcorr} the values of the observed \oiii\ luminosities, $\log$ $L^{\rm obs}_{\oiii}$, correspond to those used in PNLF studies.  
Most galaxies have their upper values of $\log$ $L^{\rm obs}_{\oiii}$ at $2.54\pm0.25$, with a few exceptions (NGC\,3109, NGC\,147). Note that, although spectroscopic studies have been made on bright PNe,  PNLF samples may contain a larger number of bright PNe than those reported here, so that the luminosity distribution of the PNe of our samples is not necessarily identical with the PNLF derived from photometric observations of large samples of PNe. Note also that the distances used in this paper are those recommended by \citet{jarrett19} based on an analysis of several distance indicators, and, in some cases, are somewhat distinct from the PNLF distances.

\section{Conclusions}
\label{conclusions}

With the aim of a comparative study of the PN populations in different galaxies, we have collected a sample of published spectroscopic data on 470 objects in 13 galaxies. The data were selected according to various requirements so that the final sample is as homogeneous as possible. The star formation histories and morphological types of the galaxies used here cover a wide range and thus, their PN progenitor stars are expected to have different masses.

We are aware that the samples of PNe studied here are not complete. An adequate analysis of the populations of PNe in external galaxies requires to have data of all the PNe located in a particular field of each of these galaxies. We also want to highlight that by summarizing all the available observations of extragalactic PNe that fall within a set of selection criteria, we were able to obtain some interesting findings, even though the samples are not complete or homogeneous. 

The emission lines compiled are 28 but in this paper we studied only a few of them. The others will be used in a future paper to study chemical abundances. If upper limits of [\ion{O}{ii}] $\lambda$3727, [\ion{O}{iii}] $\lambda$5007, \ion{He}{i} $\lambda$5876, and \heii\ $\lambda$4686 are not provided by the authors, we have adopted as an upper limit the uncertainty of a nearby and weak line. A few references do not provide uncertainties for the line fluxes and we estimated them from a logarithmic linear fit to the relative uncertainties vs. line fluxes from other observations with similar characteristics. We have also compiled the available [\ion{O}{iii}] $\lambda$5007 magnitudes to derive [\ion{O}{iii}] luminosities. If the magnitudes were not available we have used the spectroscopic data to compute the luminosities. The comparison between the theoretical and observed values of the [\ion{N}{ii}] $\lambda$6584/$\lambda$6548, [\ion{O}{iii}] $\lambda$5007/$\lambda$4959, and H$\gamma$/\Hb ratios allow us to find out if the uncertainties were adequately estimated, if there could be spectral calibration problems, or if the stellar absorption has not been taken into account correctly.  

We have studied the electron densities derived from \sii\ and \ariv\ lines. Most of the PNe have densities in the range 1\,000--10\,000 cm$^{-3}$. In the sample studied here there is a general tendency for the \ariv\ density to be higher than the \sii\ density, in some cases by a large factor. This result is in contradiction with recent findings by \citet{rodriguez20} in Galactic PNe. The fact that the observed portion in Galactic are generally much smaller than the entire objects could perhaps explain the difference, but this hypothesis needs testing. 

A crucial aspect in the study of extragalactic PNe is the discrimination between PNe and other objects such as \ion{H}{ii} regions and SNR. We have explored several diagnostics that help us to identify PNe. The ionized masses and \Hb luminosities reveal that around 30 objects identified as PNe in previous papers are likely compact \ion{H}{ii} regions. In particular this concerns most of the objects in M\,81 in the sample of \citet{stanghellini10}. On the other hand, we found that the five objects from \citet{pena12} classified as compact \ion{H}{ii} regions by \citet{roth18} are actually PNe. From the analysis of the values of \sii/\Ha and \Hb luminosities we concluded that there are not supernova remnants in our sample.

We found a clear anticorrelation between the ionized masses and the electron densities in M\,31, M\,33, and NGC\,300 that suggests that most of the PNe are ionization bounded. This trend is completely absent in LMC and SMC which could be indicating that many of the PNe observed in these galaxies are density bounded.

We found no correlation between the electron densities and the \Hb luminosities which indicates that the brightest PNe are not necessarily the youngest. An increase of \cHb\ with the \Hb luminosities is found in M\,31, M\,33, NGC~300, LMC, and NGC\,6822 which can be caused by a higher amount of dust in more massive progenitor stars as suggested by \citet{ciardullo99}. These two results that are related to the progenitor masses will be studied in a future paper.

Finally we have studied the correlations between \Heii/\Hb, $\log$ \Oiii/\Oii, and \LHb\ which are not directly dependent on abundances. The PNe in LMC, M\,33, and possibly SMC have higher values of \Heii/\Hb than the PNe in other galaxies. According to photoionization models, this indicates that many PNe in LMC and some in M\,33 and SMC are density bounded, in agreement with the behaviour of the ionized masses with respect to the electron densities. The PNe studied here cover a wide range in the \Oiii/\Oii\ values being M\,33 the galaxy where the highest values are found and the early type galaxies those with the lowest values. 

As for the \LHb--\Loiii\ relation we found that they are strongly correlated. This means that the upper limit of the PNLF based on \oiii\ magnitudes and used as a  means to measure galaxy distances, can be investigated by first analyzing \Hb (or \Ha) luminosities, which are much easier to interpret in terms of the the populations of PN progenitors. The difference between \LHb--\Loiii\ is a second order effect on the \oiii\ PNLF which can be studied in a second step.

In conclusion, the relatively simple analysis performed here from a few intensity ratios and parameters derived from these intensity ratios allows us to obtain some first findings. First, we show that there are a few parameters that are essential to distinguish PNe from other related objects: the ionized mass and the \LHb\ to discriminate between PNe and compact \hii regions and the \sii/\Ha and the \LHb\ to identify SNRs. Second, we found some indications that the PNe (and their progenitors) are different in the different galaxies: 1) most of the PNe in the spiral galaxies M\,31, M\,33, and NGC\,300 seem to be ionization bounded whereas many of those in the irregular galaxies LMC and SMC are likely density bounded, 2) The extinction appears to be correlated with the intrinsic luminosity, which is an important clue for the study of the PN luminosity function based on \oiii\ magnitudes. Further papers of this series will analyze topics related the PN progenitors and their evolution as well as topics related to their chemical composition.

\section*{ACKNOWLEDGMENTS}
We are very grateful to the anonymous referee for his/her insightful comments that have contributed to improve the final version of the paper. The authors acknowledge C. Morisset for his constant support with PyNeb and PyCloudy.
GD-I and GS acknowledge support from DGAPA-PAPIIT grants IA-101517 and IN-103820. JG-R acknowledges support from an Advanced Fellowship from the Severo Ochoa excellence program (SEV-2015-0548) and from the State Research Agency (AEI) of the Spanish Ministry of Science, Innovation and Universities (MCIU) and the European Regional Development Fund (FEDER) under grant AYA2017-83383-P. JG-R also acknowledges support under grant P/308614 financed by funds transferred from the Spanish Ministry of Science, Innovation and Universities, charged to the General State Budgets and with funds transferred from the General Budgets of the Autonomous Community of the Canary Islands by the MCIU. GD-I and JSR-G acknowledges CONACyT grant 241732. JSR-G also acknowledges the postdoctoral fellowship funded by Direcci\'on General de Asuntos del Personal Acad\'emico (DGAPA), Universidad Nacional Aut\'onoma de M\'exico (UNAM). GS, JG-R, and JSR-G are grateful to G. Delgado Inglada and A. Farah for having invited them to stay in their home where a large part of this work was carried out. We thank Fabio Herpich for help with the WISE data. This research has made use of the NASA/IPAC Extragalactic Database (NED), which is funded by the National Aeronautics and Space Administration and operated by the California Institute of Technology.".

\section*{Data availability}
This paper use data compiled from published sources whose references are given in the text. The whole compilation will be made publicly available when all the papers of the series are published.

\bibliographystyle{mnras}
\bibliography{literature}

\FloatBarrier
\appendix
\section{Maximum values of \heii/H$\beta$ in PNe}\label{app1}
Under the condition that the nebular gas absorbs all the ionizing photons from the central star, the ratio \heii/\Hb is  an indicator of the effective temperature. However, as shown by \citet{stasinska86}, the \heii/\Hb ratio also depends on the ionization parameter. The reason is that not all photons with energies above 54.4 eV are absorbed by \Hep, some are absorbed by neutral hydrogen atoms, whose abundance with respect to \Hep\ increases in the \Hepp\ zone as the ionization parameter decreases. In addition, the \heii/\Hb ratio levels off at temperatures above $250-350$ kK, because the proportion of \Hep\ ionizing photons emitted by the star increases more slowly with temperature.

To illustrate this and find the maximum value of \heii/H$\beta$ expected in ionization bounded PNe, we have computed two sequences of ionization bounded models using Cloudy \citep{Ferland17} version 17.01 and {\sc PyCloudy} \citep{Morisset13,Morisset14}. The exciting star is a blackbody, with effective temperature going 
from 50,000 to 350,000 K, the log of the mean ionization parameter, <U>, ranges from $-4$ to $-2$, the electron density is $10^3$ cm$^{-3}$, and the geometry is a filled sphere. The abundances of the heavy elements are the solar photospheric ones from \citet{asplund09}. 

\begin{figure}
  \centering
  \includegraphics[width=4cm, trim=0 0 0 0]{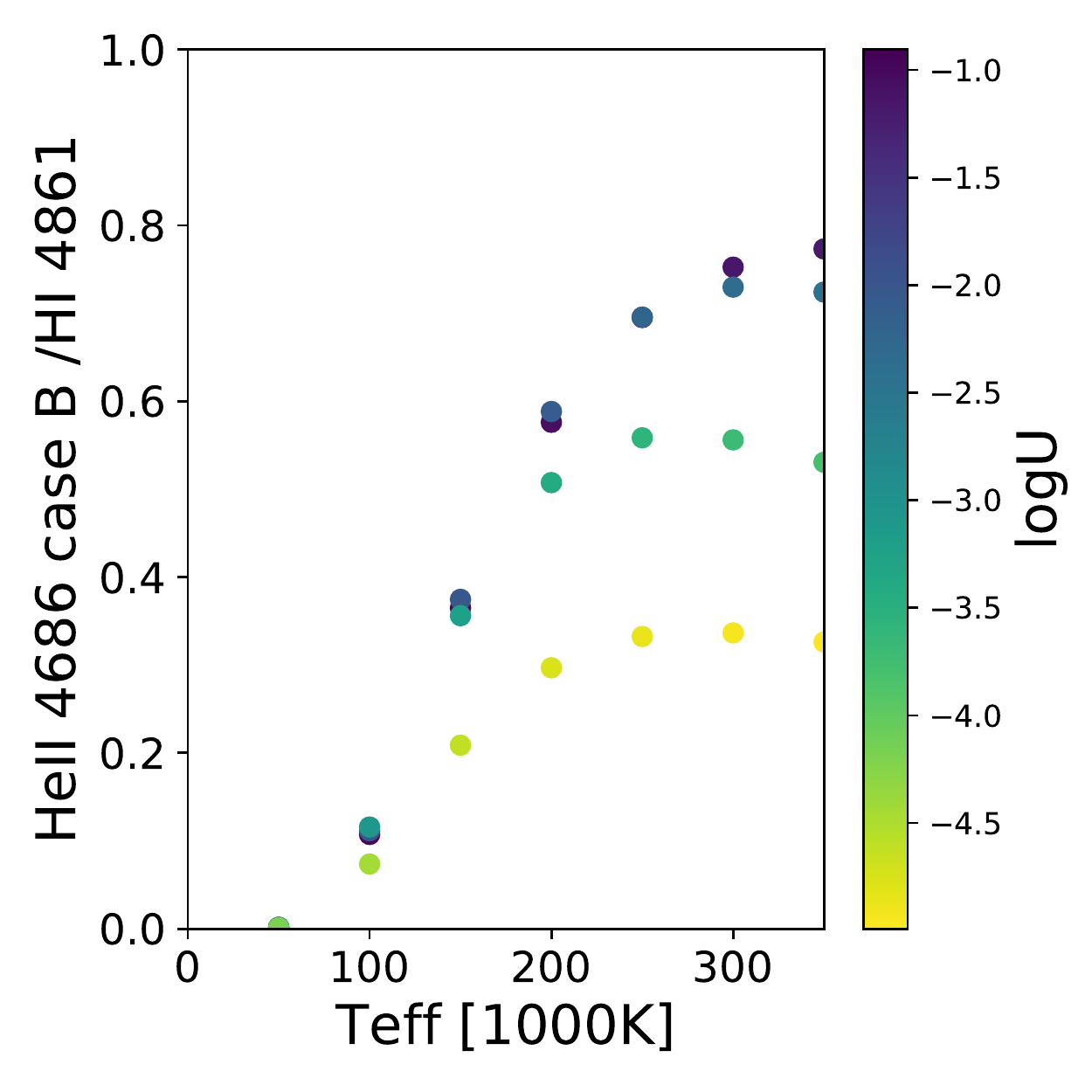}
    \includegraphics[width=4cm, trim=0 0 0 0]{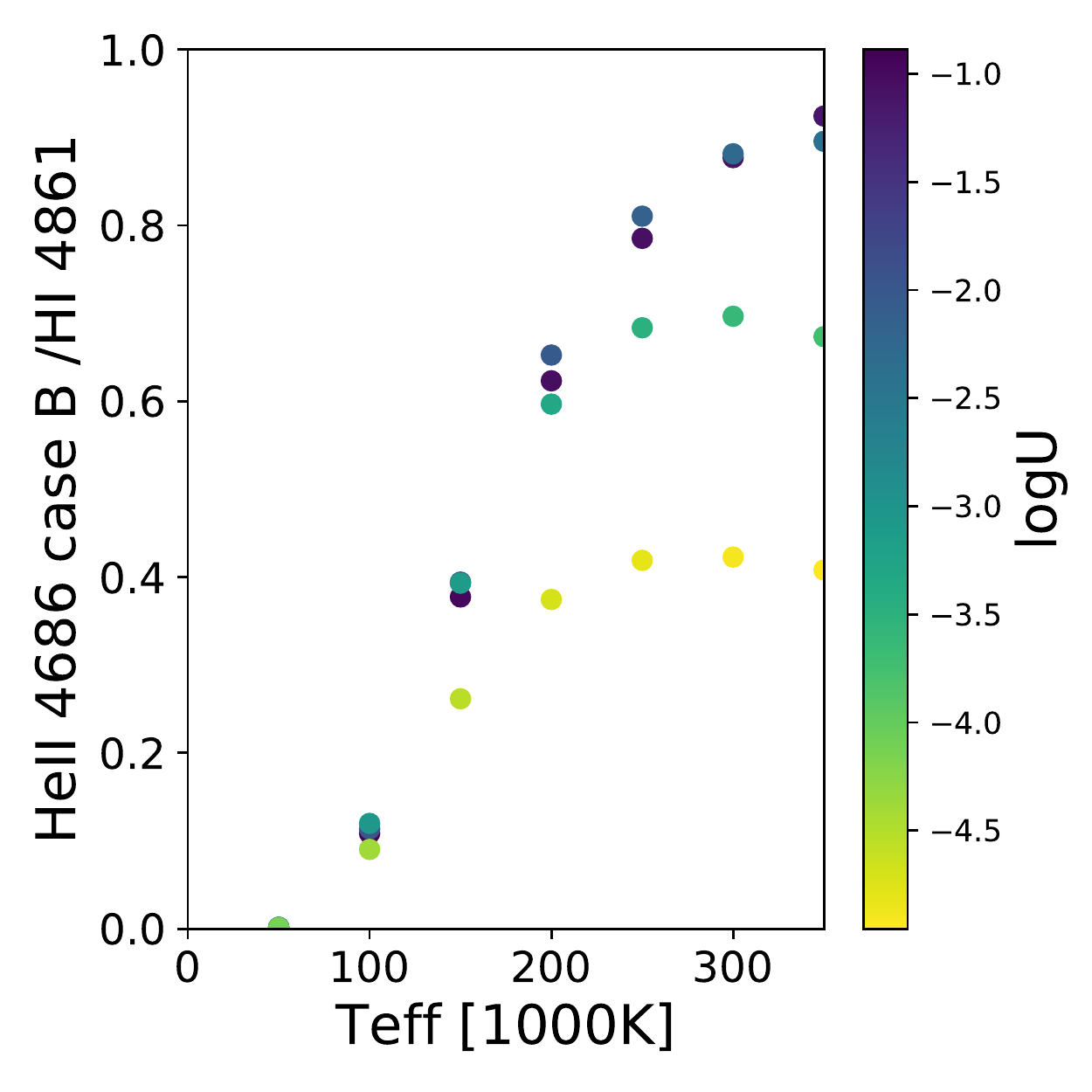}
  \caption{Values of \heii/\Hb as a function of the effective temperature for a sample of photoionization models. Left panel:  He/H$=0.10$, right panel He/H$=0.15$. The colorbar runs from low to high ionization parameter values.}
  \label{fig:models}
\end{figure}

Figure~\ref{fig:models} shows the resulting values of {\heii} (case B)/\Hb as a function of the effective temperature for the computed models\footnote{We plot the case B values, because our models considered a small number of n-resolved levels, and case B gives an upper limit to  the real {\Heii} emissivities.}. In the left panel the models have  He/H$=0.10$, in the right one they have  He/H$=0.15$. We see that the maximum value of \heii/\Hb reached by the models is $\sim0.9$.   

However, according to the evolution models for post-AGB stars by \citet{millerbertolami16}, the effective temperature exceeds 250 kK only for the most massive stars and for a very short time, so such cases are expected to be rare.

\section{Distinguishing planetary nebulae from supernova remnants}\label{app2}

\begin{figure}
  \centering
  \includegraphics[width=\columnwidth, trim= 20 0 40 0cm]{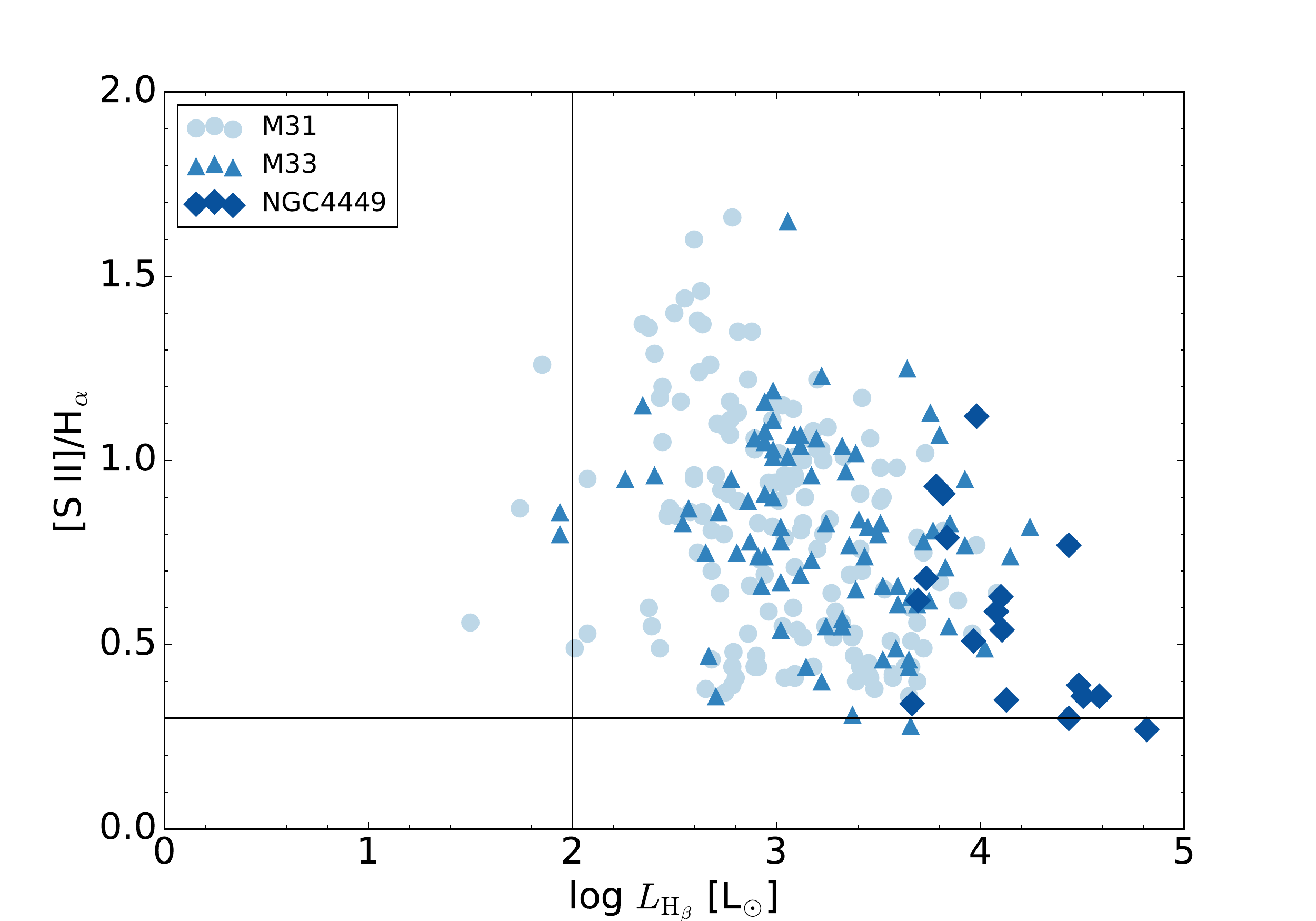}
  \caption{Values of \sii/\Ha as a function of $\log$\LHb\ for supernova remnants in M\,31, M\,33, and NGC\,4449.}
  \label{fig:snr}
\end{figure}

We have compiled the values of \ion{S}{ii}/H$\alpha$ and H$\alpha$ luminosities for supernova remnants in M\,31, M\,33, and NGC\,4449 from \citet{Lee14, Long2010} and \citet{Leonidaki13}, respectively. The values are plotted in Figure~\ref{fig:snr}. The lower value of \ion{S}{ii}/H$\alpha$ in this sample of SNR is $\sim0.3$. On the other hand, $\log$\LHb$\sim2$ can be taken as the lower value of the {\Hb} luminosity  since only 5 out of 560 SNRs are below this limit.

\label{lastpage}
\end{document}